\documentclass[pre,twocolumn]{revtex4}

\usepackage{natbib}
\usepackage{amssymb,amsmath}
\usepackage{epsfig}
\usepackage{bm}
\usepackage{graphicx}
\usepackage{hyperref}

\usepackage{graphicx}

\usepackage{color}

\newcommand\beq{\begin{equation}}
\newcommand\eeq{\end{equation}}
\newcommand\beqa{\begin{eqnarray}}
\newcommand\eeqa{\end{eqnarray}}
\newcommand{\dd}{\text{d}}

\newcommand{\al}{\alpha}

\begin{document}

\title{Non-Newtonian hydrodynamics for a dilute granular suspension under uniform shear flow}

\author{Mois\'es G. Chamorro\footnote[1]{Electronic address:moises@unex.es}}
\affiliation{Departamento de F\'{\i}sica and Instituto de Computaci\'on Cient\'{\i}fica Avanzada (ICCAEx),
Universidad de Extremadura, E-06071 Badajoz, Spain}
\author{Francisco Vega Reyes\footnote[2]{Electronic address:fvega@unex.es;
URL: http://http://www.unex.es/eweb/fisteor/fran/}}
\author{Vicente Garz\'{o}\footnote[3]{Electronic address: vicenteg@unex.es;
URL: http://www.unex.es/eweb/fisteor/vicente/}}
\affiliation{Departamento de F\'{\i}sica and Instituto de Computaci\'on Cient\'{\i}fica Avanzada (ICCAEx), Universidad de Extremadura, E-06071 Badajoz, Spain}

\begin{abstract}

We study in this work a steady shearing laminar flow with null heat flux (usually called ``uniform shear flow'') in a
gas-solid suspension at low density.  The solid particles are modeled as a gas of smooth hard spheres with inelastic collisions while the influence of the surrounding interstitial fluid on the dynamics of grains is modeled by means of a volume drag force, in the context of a rheological model for suspensions. The model is solved by means of three different but complementary routes, two of them being theoretical (Grad's moment method applied to the corresponding Boltzmann equation and an exact solution of a kinetic model adapted to granular suspensions) and the other being computational (Monte Carlo simulations of the Boltzmann equation). Unlike in previous studies on granular sheared suspensions, the collisional moment associated with the momentum transfer is determined in Grad's solution by including \emph{all} the quadratic terms in the stress tensor. This theoretical enhancement allows us for the detection and evaluation of the normal stress differences in the plane normal to the laminar flow. In addition, the exact solution of the kinetic model gives the explicit form of the velocity moments of the velocity distribution function. Comparison between our theoretical and numerical results shows in general a good agreement for the non-Newtonian rheological properties, the kurtosis (fourth velocity moment of the distribution function) and the velocity distribution of the kinetic model for quite strong inelasticity and not too large values of the (scaled) friction coefficient characterizing the viscous drag force. This shows the accuracy of our analytical results that allows us to describe in detail the flow dynamics of the granular sheared suspension.

\end{abstract}

\draft
\date{\today}
\maketitle

\section{Introduction}
\label{sec1}

The study of granular matter is of interest in a wide variety of fields in fundamental and applied science: different industry and technology sectors, biophysics, fluid mechanics, statistical physics, and even in optics applications. As a consequence, there is a large bibliography on granular dynamics. As it is known, and depending on the particle density of the granular system, its dynamics and in consequence, its theoretical modeling, can be very different \cite{G03}.

Generically, we may differentiate the high and low density regimes, where the latter is essentially characterized by binary particle collisions and the former presents multiparticle collisions/contacts. We will focus on the binary collision regime where  the system is usually called a ``granular gas.'' Since particle collisions are inelastic by definition, a direct consequence is that the low density regime can only be maintained if there is some kind of energy input in the system. Otherwise, if the granular gas is left to freely cooling, it will eventually collapse by a mechanism of clustering instabilities \cite{MIMA08} (that is increasingly stronger with increasing inelasticity) \cite{GZ93,M93,WK00}.

On the other hand, although in nature granular particles are frequently surrounded by an interstitial fluid (like the air, for instance), the influence of the latter on the dynamic properties of solid particles is generally neglected in most theoretical and computational works. However, the effect of the interstitial fluid on solid particles turns out to be significant in a wide range of practical applications and physical phenomena \cite{IH95}, like for instance species segregation (see for instance, Refs.\ \cite{MLNJ01,NSK03,Yan03,SSk04,MCEKNJ05,WZXS08,ZHK08,Idler09,CPSK10,PGM14}) or in biophysics where active matter may be considered as a driven granular suspension \cite{KSRS14}. For this reason the study of gas-solid flows has attracted the attention of engineering and the physics community in the last few years \cite{KH01}.

The description of gas--solid suspensions, whose dynamics is very complex, is a long-standing branch of classic fluid mechanics \cite{B74}. For instance, particles suspended in a fluid feel a lubrication force, transmitted by the surrounding fluid but originated by the presence of another nearby particle. It is known that this kind of interaction (usually called ``hydrodynamic interaction'') depends also on the global configuration of the set of grains \cite{BB88}, giving rise to tensor-rank force equations. The modeling of these lubrication forces is rather involved and several approaches can be used. For this reason, there is a large bibliography, that extends for decades and that is devoted to the study of this kind of interactions (Stokesian or Stokes dynamics) \cite{BB88,BGP11,LXWZJ14}. Nevertheless, in the dilute suspension limit, these hydrodynamic interactions become less relevant \cite{BB88,B74} and only the isolate body resistance is retained, usually in the form of a simple drag force. On the other hand, due to the inherent complexity of the interaction between the interstitial fluid and the granular particles, early kinetic theory studies have neglected in most cases the effect of inelasticity in suspended particle collisions \cite{TMS84,LMJ91,MS94,SM94,TK95}. This kind of approach is not entirely accurate since of course in most real cases the sizes of suspended particles are big enough to render particle collisions inelastic (bigger than $1~\mu\mathrm{m}$, otherwise particles may be considered as colloids, for which collisions are elastic \cite{LXWZJ14,GHP15}). Therefore, inelasticity in the collisions can play a major role in the dynamics of granular (as opposed to colloidal) suspensions, specially in the dilute limit at high Stokes number, where grain-grain collisions effects dominate over many particle hydrodynamic interactions \cite{E06}. However, only more recent works have dealt with inelastic collisions in the case of dilute \cite{YPTKT01,TFAJ07} and moderately dense \cite{SMTK96} suspensions.

At a kinetic theory level, the description of granular suspensions is an intricate problem since it involves two phases (solid particles and interstitial fluid) and hence, one would need to solve a set of two coupled kinetic equations for each one of the velocity distribution functions of the different phases. However, due to the mathematical difficulties embodied in this approach and in order to gain some insight into this problem, a usual model \cite{KH01,K90} for gas-solid flows is to consider a single Boltzmann equation for the solid particles \cite{BP04} where the influence of the surrounding fluid on them is modeled by means of an effective external force. This will be the approach considered in the present paper.

Moreover, in the study of granular suspensions usually only simple states have been considered, due to the inherent complexity of the system. For instance, in a recent work \cite{GTSH12} the Navier-Stokes transport coefficients of monodisperse gas-solid flows at \emph{moderate} densities were obtained by solving a model based on the Enskog kinetic equation by means of the application of the Chapman-Enskog method \cite{CC70} around the so-called homogeneous cooling state (HCS). The external force $\mathbf{F}_\text{ext}$ proposed in Ref.\ \cite{GTSH12} to model the effect of the fluid phase on grains is composed by three different terms: (i) a term proportional to the difference between the mean flow velocities of solid $\mathbf{U}$ and gas $\mathbf{U}_g$ phases, (ii) a drag force $\mathbf{F}_\text{drag}$ proportional to the velocity of particle and (iii) a stochastic force $\mathbf{F}_\text{st}$ accounting for particle neighbor effects (Langevin model). In the case that $\mathbf{U}=\mathbf{U}_g$, the coefficient associated with the stochastic force vanishes and only the drag force interaction $\mathbf{F}_\text{drag}$ remains, namely, mean drag and neighbor effects disappear in the suspension model of Ref.\ \cite{GTSH12}. It is important to remark that the above drag force model has been also recently considered in different papers \cite{WZLH09,H13,HT13,SMMD13,WGZS14} to study the shear rheology of frictional hard-sphere suspensions.

Nevertheless, the ranges of interest of the physics of granular gases fall frequently beyond Newtonian hydrodynamics since the strength of the spatial gradients is large in most situations of practical interest (for example, in steady states). This is essentially due to the coupling between collisional dissipation and spatial gradients that under steady states usually yields moderately large spatial gradients \cite{G03,SGD04,VU09}. In these steady states, a hydrodynamic description is still valid but with constitutive equations more complex than the Navier-Stokes ones \cite{VSG10,VSG13}. A very neat example of this is the simple or uniform shear flow (USF) \cite{C90}, that except in the quasi-elastic limit, is essentially non-newtonian \cite{L04,Ku04,L06,G06,K06,SA14}. It is characterized by a linear velocity field (that is $\partial U_x/\partial y\equiv a =\text{const}$), constant density $n$ and constant temperature $T$. In particular, in the USF state the presence of shearing induces anisotropies in the pressure tensor $P_{ij}$, namely, nonzero shear stress $P_{xy}$ and normal stress differences $P_{xx}-P_{yy}$ and $P_{yy}-P_{zz}$. In addition, in the case of granular suspensions, it may be assumed \cite{TK95,SMTK96} that $\mathbf{U}=\mathbf{U}_g$ and so, $\mathbf{F}_\text{ext}=\mathbf{F}_\text{drag}$. Here, the number density $n$, the mean flow velocity $\mathbf{U}$ and the granular temperature $T$ are defined, respectively, as
\beq
\label{2.3}
n(\mathbf{r},t)=\int\; \dd \mathbf{v}\; f(\mathbf{r},\mathbf{v},t),
\eeq
\beq
\label{U}
\mathbf{U}(\mathbf{r},t)=\frac{1}{n(\mathbf{r},t)}\int\; \dd \mathbf{v}\; \mathbf{v} f(\mathbf{r},\mathbf{v},t),
\eeq
\beq
\label{2.5.1}
T(\mathbf{r},t)=\frac{2}{d n(\mathbf{r},t)}\int\; \dd \mathbf{v} V^2 \; f(\mathbf{r},\mathbf{v},t).
\eeq
where $f(\mathbf{r},\mathbf{v},t)$ is the one-particle velocity distribution function and $\mathbf{V}=\mathbf{v}-\mathbf{U}$ is the peculiar velocity.



A detailed study of simple shear flows of granular suspensions at finite Stokes numbers was carried out  by Tsao and Koch \cite{TK95} and Sangani \emph{et al.} \cite{SMTK96}. In both of these works, and like in the model used in Ref.\ \cite{GTSH12},  suspension dynamics is dominated by the drag exerted by the fluid (external drag force) and the solid-body collisions between the particles. In the first paper \cite{TK95}, the authors considered a \emph{dilute} gas-solid suspension of \emph{elastic} particles, thus neglecting the important effect of inelasticity in macroscopic particles. Inelasticity and excluded volume effects (moderated densities) were only considered in the second paper \cite{SMTK96} of the series. Moreover, in the first reference \cite{TK95} (elastic collisions), Tsao and Koch solved the Boltzmann kinetic equation by means of a Grad's moment method approach \cite{G49} where the collisional moment $\Lambda_{ij}$ of the momentum transfer (see Eq.\ \eqref{3.14} for its definition) was evaluated by retaining \emph{all} the quadratic terms in the pressure tensor $P_{ij}$ (nonlinear Grad's solution). However, for practical applications, in their actual theoretical results only the term proportional to the shear stress $P_{xy}^2$ was retained in the nonlinear contributions to $\Lambda_{ij}$, see Eqs.\ (3.14a,b) of \cite{TK95}. Sangani \emph{et al.} \cite{SMTK96} solved first the Enskog kinetic equation (which is an extension of the Boltzmann equation to dense systems) by means of Grad's method but only \emph{linear} terms in the shear rate and the pressure tensor (linear Grad's solution) were retained in their calculation of $\Lambda_{ij}$ (see Eq.\ (4.21) of \cite{SMTK96}). Some discrepancies were observed in the very dilute regime for the normal stress differences. In particular, their linear Grad's solution yields $P_{yy}=P_{zz}$ (see Eq.\ (4.33) of \cite{SMTK96}) which clearly disagrees with simulation results \cite{SMTK96}.


The objective of this paper is to offer a complete study of the USF state for \emph{dilute} granular suspensions where the effect of fluid phase on grains is taken into account by the presence of an external drag force in the kinetic equation. For the accomplishment of this task, we propose in this work three different approaches: two of them are theoretical and the third one is computational. In the first theoretical approach, the Boltzmann equation is solved by Grad's method where both inelasticity and at the same time all of the non-linear terms in shear rate and stress tensor are retained in our expression of the collisional moment $\Lambda_{ij}$. Thus, as we will see, new interesting properties of the suspension arise from this refinement.  For instance, we are able to detect the influence of both viscous friction and inelasticity on the normal stress difference $P_{yy}-P_{zz}$. In this sense, our theory  generalizes previous analyses \cite{TK95,SMTK96}, these being  recovered when the appropriate simplifications are applied to our theory.

Apart from Grad's method, we also use a second theoretical approach based on the derivation of an \emph{exact} solution to a simplified model kinetic equation \cite{BDS99} for the sheared granular suspension. This will allow us to determine all the velocity moments of the velocity distribution function as well as the explicit form of the latter in terms of the shear rate $a$, the friction coefficient $\gamma$ characterizing the drag force and the coefficient of restitution $\alpha$. In particular, the rheological properties derived from the BGK solution are the same as those obtained in linear Grad's solution to the Boltzmann equation.

As a third route and to gauge the accuracy of the previous analytical results, we numerically solve the Boltzmann equation for the granular suspension by means of the direct simulation Monte Carlo (DSMC) method \cite{B94}. This (exact) numerical solution takes into account the real grain-grain collisions in the context of hard sphere collision model. As we will see, the comparison between theory and simulation shows that both (approximate) solutions give in general accurate results even for conditions of quite strong inelasticity (say for instance, $\alpha \gtrsim 0.5$). Moreover, the theoretical predictions for $P_{yy}$ and $P_{zz}$ obtained from our nonlinear Grad's solution agree very well with simulations (see Fig.\ \ref{figPii}), showing the improvement of our theory with respect to the previous analysis of Sangani \emph{et al.} \cite{SMTK96}. On the other hand, the agreement between theory and simulation become worse as the (scaled) friction coefficient $\gamma^*$ increases. This means that our theory of rapidly sheared granular flows become more reliable as the effects of the inelastic particle collisions dominate over viscous effects.

The plan of the paper is as follows. In Sec.\ \ref{sec2}, the Boltzmann equation of inelastic hard spheres driven by an external drag force is introduced and the USF problem for granular suspensions is presented. The analytical results derived in the paper are provided in Sec. \ref{sec3} whereas some technical details on the DSMC method used here are briefly described in Sec.\ \ref{sec4}. Section \ref{sec5} deals with the comparison between theory and simulation results. Finally, the paper is closed in Sec. \ref{sec5} with a brief discussion on the results reported in the present contribution.

\section{Description of the system}
\label{sec2}

\subsection{Boltzmann kinetic equation for granular suspensions}

	 Let us consider a set of solid particles of mass $m$ and diameter $\sigma$ immersed in a viscous gas. As we already commented, for big enough particles (typical size $\lesssim 1~\mu\mathrm{m}$), collisions between particles carry a partial loss of their kinetic energy. Thus, the solid particles can be modeled as a gas of smooth hard spheres (or disks, for two-dimensional systems) with inelastic collisions. The inelasticity of collisions is characterized by a (positive) \emph{constant} coefficient of normal restitution $0\leq \al \leq 1$, where $\alpha=1$ stands for completely elastic collisions and $\alpha=0$ for completely inelastic collisions \cite{SG98,BDKS98,G03}.

	In the dilute limit, the corresponding Langevin equation describing the gas–-solid interaction force can be greatly simplified \cite{B74,B74b}. There are several experimental results on the dynamics of dilute particle systems immersed in a gas flow that validate this kind of approach. For instance, this type of system was analyzed in early experimental studies where the corresponding flow properties were carefully measured \cite{TMS84}. These experimental results were later used for validation of a hydrodynamic theory of a granular suspension immersed in gas flow, allowing for characterization of the relevance of grains collisions in the hydrodynamic behavior of the turbulent suspension \cite{LMJ91}. It has been shown more recently, in experiments, that the turbulent gas-grain interaction can also be described by a Langevin equation with a stochastic force that has the form of a white noise, much in the same way as in classic studies at lower Reynolds number \cite{BB88}. Therefore, under the above conditions one can consider the following generalized Langevin model for the instantaneous acceleration on a suspended grain:
\begin{equation}
		 m\frac{\mathrm{d}\mathbf{v}}{dt}=-\beta(\mathbf{U}-\mathbf{U}_g)-\gamma\cdot\mathbf{V}+\mathbf{F}_{\mathrm{st}},
		\label{langevin}
\end{equation}
where $\mathbf{F}_\mathrm{st}$ is a stochastic force with the following properties \cite{BB88}
\begin{equation}
\langle {\bf F}_i^{\text{st}}(t) \rangle ={\bf 0}, \quad \langle {\bf F}_i^{\text{st}}(t) {\bf F}_j^{\text{st}}(t') \rangle =\openone m^2 \xi \delta_{ij}\delta(t-t').
\label{Fst}
\end{equation}
In Eq.\ \eqref{Fst}, $\openone$ is the $d\times d$ unit matrix and $\xi$ represents the strength of the correlation.
The model described by Eq.\ \eqref{langevin} has been recently proposed in Ref.\ \cite{GTSH12} for monodisperse gas-solid flows at moderate density. Although the coefficients $\beta$, $\gamma$, and $\xi$ appearing in Eqs.\ \eqref{langevin} and \eqref{Fst}, respectively, are in general tensors, in the case of a dilute suspension they may be simplified as scalars \cite{BB88}. Those coefficients are associated with the instantaneous gas-solid force \cite{GTSH12}. As we said in the Introduction, the first term on the right-hand side of Eq.\ \eqref{langevin} represents the portion of the drag term arising from the mean motion of particle and solid phase; the second term is traced to fluctuations in particle velocity (relative to its mean value) and finally the third term is a stochastic model for the change in particle momentum due to shear stress and pressure contributions at the particle surface that arise from the fluid velocity and pressure disturbances caused by neighbor particles.

According to the model proposed in Ref.\ \cite{GTSH12}, at low mean Reynolds number, the expressions of $\gamma$ and $\xi$ for dilute suspensions of hard spheres are, respectively, \cite{GTSH12}
\begin{equation}
\label{2.4}
\gamma=\frac{m}{\tau}R_\mathrm{diss}(\phi),
\end{equation}
\begin{equation}
\label{2.5}
\xi=\frac{1}{6\sqrt{\pi}}\frac{\sigma |\Delta {\bf U}|^2}{\tau^2
\sqrt{\frac{T}{m}}},
\end{equation}
where $\tau=m/(3\pi \mu_g \sigma)$ is the characteristic time scale over which the velocity of a particle of mass $m$ and diameter $\sigma$ relaxes due to viscous forces, $\mu_g$ being the gas viscosity. Moreover,  $\phi=(\pi/6)n\sigma^3$ is the solid volume fraction for spheres,
\beq
\label{Rdiss}
R_\text{diss}(\phi)=1+3\sqrt\frac{\phi}{2},
\eeq
and $\Delta \mathbf{U}=\mathbf{U}-\mathbf{U}_g$.

In the low-density regime the one-particle particle distribution function $f(\mathbf{r}, \mathbf{v},t)$ provides complete information on the state of the system. This quantity gives the average number of particles that at instant $t$ are located around the point $\mathbf{r}$ and with a velocity about $\mathbf{v}$. In the case of an external force composed by the three terms appearing in Eq.\ \eqref{langevin}, the corresponding Boltzmann kinetic equation for dilute granular suspensions is \cite{GTSH12}
\begin{align}
& \partial_{t}f+\mathbf{v}\cdot \mathbf{\nabla}f-\frac{\beta}{m}\Delta {\bf U}\cdot
\frac{\partial f}{\partial {\bf V}}-\frac{\gamma}{m} \frac{\partial}{\partial
{\bf V}}\cdot {\bf V} f \nonumber \\
&-\frac{1}{2}\xi\frac{\partial^2}{\partial V^2}f=J\left[f,f\right],
\label{2.6}
\end{align}
where the Boltzmann collision operator $J\left[{\bf v}|f,f\right]$ is given by
\beqa
\label{2.7}
J\left[{\bf v}_{1}|f,f\right]&=&\sigma^{d-1}\int \dd {\bf v}
_{2}\int \dd \widehat{\boldsymbol{\sigma}}\,\Theta (\widehat{{\boldsymbol {\sigma }}}
\cdot {\bf g})(\widehat{\boldsymbol {\sigma }}\cdot {\bf g})\nonumber\\
& & \times \left[\alpha^{-2}f({\bf v}_1')f({\bf v}_2')-f({\bf v}_1)f({\bf v}_2)\right].\nonumber\\
\eeqa
Here, $d$ is the dimensionality of the system ($d=2$ for disks and $d=3$ for spheres), $\boldsymbol
{\sigma}=\sigma \widehat{\boldsymbol {\sigma}}$, $\widehat{\boldsymbol {\sigma}}$ being a unit vector pointing in the direction from the center of particle $1$ to the center of particle $2$, $\Theta $ is the Heaviside step function, and ${\bf g}={\bf v}_{1}-{\bf v}_{2}$ is the relative velocity. The primes on the velocities in Eq.\ \eqref{2.7} denote the initial
values $\{{\bf v}_{1}^{\prime}, {\bf v}_{2}^{\prime }\}$ that lead to
$\{{\bf v}_{1},{\bf v}_{2}\}$ following a binary collision:
\begin{subequations}
\begin{equation}
{\bf v}_{1}^{\prime}={\bf v}_{1}-\frac{1}{2}\left( 1+\alpha^{-1}\right)
(\widehat{{\boldsymbol {\sigma }}}\cdot {\bf g})\widehat{{\boldsymbol {\sigma }}},
\label{2.8}
\end{equation}
\begin{equation}
{\bf v}_{2}^{\prime}={\bf v}_{2}+\frac{1}{2}\left( 1+\alpha^{-1}\right)
(\widehat{{\boldsymbol {\sigma }}}\cdot {\bf g})\widehat{{\boldsymbol {\sigma }}}.
\label{2.8.1}
\end{equation}
\end{subequations}

From the Boltzmann equation \eqref{2.6} one can derive the (macroscopic) hydrodynamic equations for the number density $n$ , the flow velocity $\mathbf{U}$ and the granular temperature $T$. They are given by
\beq
\label{2.9}
D_t n+n \nabla \cdot \mathbf{U}=0,
\eeq
\beq
\label{2.10}
D_t \mathbf{U}+(m n)^{-1} \nabla \cdot \mathsf{P}=-\frac{\beta}{m}\Delta \mathbf{U},
\eeq
\beq
\label{2.11}
D_tT+\frac{2}{d n}\left( \nabla \cdot \mathbf{q}+\mathbf{P}:\nabla \mathbf{U} \right)=-\frac{2T}{m}\gamma+m\xi-T \zeta.
\eeq
Here, $D_t\equiv \partial_t+\mathbf{v}\cdot \nabla$ is the material derivative,
\beq
\label{2.12}
P_{ij}=m\int\; \dd \mathbf{v} V_i V_j f(\mathbf{v}),
\eeq
is the pressure tensor,
\beq
\label{2.13}
\mathbf{q}=\frac{m}{2}\int\; \dd \mathbf{v} V^2 \mathbf{V} f(\mathbf{v}),
\eeq
is the heat flux, and
\beq
\label{cooling_int}
\zeta=-\frac{m}{d n T}\int\; \dd \mathbf{v} V^2 J[\mathbf{v}|f,f]
\eeq
is the cooling rate characterizing the rate of energy dissipated due to collisions \cite{VSG13}.

Note that in the suspension model defined by Eqs.\ \eqref{2.6} and \eqref{2.7}, the form of the Boltzmann collision operator
$J[f,f]$ is the same as for a dry granular gas and hence, the collision dynamics does not contain any gas-phase parameter. As has been previously discussed in several papers \cite{TK95,SMTK96,WKL03}, the above assumption requires that the mean-free time between collisions is much less than the time taken by the fluid forces (viscous relaxation time) to significantly affect the motion of solid particles. Thus, the suspension model \eqref{2.6} is expected to describe situations where the stresses exerted by the interstitial fluid on particles are sufficiently small that they have a weak influence on the dynamics of grains. However, as the density of fluid increases (liquid flows), the above assumption could be not reliable and hence one should take into account the presence of fluid into the binary collisions event.

\subsection{Steady base state: the uniform shear flow}
\label{USFstate}

Let us assume now that the suspension is under \emph{steady} USF. This state is macroscopically defined by a constant density $n$ and temperature $T$ and the mean velocity $\mathbf{U}$ is
\beq
\label{2.15}
U_i=a_{ij}r_j, \quad a_{ij}=a\delta_{ix}\delta_{jy},
\eeq
where $a$ is the constant shear rate. In addition, as usual in uniform sheared suspensions \cite{TK95,SMTK96}, the average velocity of particles follows the velocity of the fluid phase and so, $\mathbf{U}=\mathbf{U}_g$. In this case, $\Delta {\bf U}=\textbf{0}$ and according to Eq.\ \eqref{2.5}, $\xi=0$. Thus, the steady Boltzmann equation \eqref{2.6} becomes
\beq
\label{2.16}
-aV_y\frac{\partial f}{\partial V_x}-\frac{\gamma}{m} \frac{\partial}{\partial
{\bf V}}\cdot {\bf V} f =J[\mathbf{V}|f,f].
\eeq
In Eq.\ \eqref{2.16} we use the USF property of spatial uniformity when the Boltzmann equation is expressed in terms of the peculiar velocity $V_i=v_i-a_{ij}r_j$ \cite{GS03}. We note that the Boltzmann equation \eqref{2.16} is equivalent to the one employed by Tsao and Koch \cite{TK95} (in the case of elastic collisions) and Sangani \emph{et al.} \cite{SMTK96}.

In the USF problem, the heat flux vanishes ($\mathbf{q}=\mathbf{0}$) and the only relevant balance equation is that of the temperature \eqref{2.11}. In the steady state and for the geometry of the USF, Eq.\ \eqref{2.11} reads
\beq
\label{2.17}
-\frac{2}{d n} P_{xy} a=\frac{2T}{m}\gamma+\zeta T.
\eeq
Equation \eqref{2.17} implies that in the steady state the viscous heating term ($-a P_{xy}>0$) is exactly compensated by the cooling terms arising from collisional dissipation ($\zeta T$) and viscous friction ($\gamma T/m$) \cite{VU09}. As a consequence, for a given shear rate $a$, the (steady) temperature $T$ is a function of the friction coefficient $\gamma$ and the coefficient of restitution $\alpha$. Note that in contrast to what happens for \emph{dry} granular gases ($\gamma=0$), a steady state is still possible for suspensions when the particle collisions are elastic ($\al=1$ and so, $\zeta=0$). Moreover, the balance equation \eqref{2.17} also holds for flows with \emph{uniform} heat flux (the so-called LTu class of Couette flows) \cite{VSG10,VGS11,VSG13} with no friction ($\gamma=0$). For this class of flows, the physical meaning of Eq.\ \eqref{2.17} is that there is an \emph{exact} balance at \emph{every} point of the system between the heating (coming from viscosity) and cooling (coming from inelasticity and friction) terms.

The USF state is in general non-Newtonian. This can be characterized by the introduction of generalized transport coefficients measuring the departure of transport coefficients from their Navier-Stokes forms. First, we define a non-Newtonian shear viscosity coefficient $\eta(a, \gamma, \al)$ by
\beq
\label{2.18}
P_{xy}=-\eta(a, \gamma, \al)a.
\eeq
In addition, while $P_{xx}=P_{yy}=P_{zz}=nT$ in the Navier-Stokes hydrodynamic order, normal stress differences are expected to appear in the USF state ($P_{xx}\neq P_{yy} \neq P_{zz})$. We are interested here in determining the (reduced) shear stress $P_{xy}^*$ and the (reduced) normal or diagonal elements $P_{xx}^*$, $P_{yy}^*$ and $P_{zz}^*$, where $P_{ij}^*\equiv P_{ij}/p$ and $p=nT$ is the hydrostatic pressure. With respect to the cooling rate $\zeta$ (which vanishes for elastic collisions \cite{TK95}), since this quantity is a scalar, its most general form is
\beq
\label{zeta}
\zeta=\zeta_0+\zeta_{2} a^2+\cdots.
\eeq
The zeroth-order contribution to the cooling rate $\zeta_0$ is \cite{NE98}
\beq
\label{2.18.1}
\zeta_0=\frac{d+2}{4d}\left(1-\alpha^2\right)\nu,
\end{equation}
where $\nu$ is an effective collision frequency of hard spheres given by
\begin{equation}
\nu=\frac{8}{d+2}\frac{\pi^{(d-1)/2}}{\Gamma(d/2)}n\sigma^{d-1}\sqrt{\frac{T}{m}}.
\label{nu}
\end{equation}
For hard spheres ($d=3$), Eq.\ \eqref{2.18.1} is consistent with the results derived for Sangani \emph{el al.} \cite{SMTK96} in the dilute limit (solid volume fraction $\phi=0$). On the other hand, given that the latter theory \cite{SMTK96} only retains linear terms in the pressure tensor in the evaluation of the collisional moment $\Lambda_{ij}$ (defined in Eq.\ \eqref{3.14}), then $\zeta_2=0$. We calculate the second-order contribution $\zeta_2$ to the cooling rate in Sec.\ \ref{sec3}. To the best of our knowledge, this contribution has not yet been computed in previous works on granular sheared suspensions.

Equation \eqref{2.17} can be rewritten in dimensionless form when one takes into account Eq.\ \eqref{2.18}:
\begin{equation}
\frac{2}{d}\eta^*a^{*2} = 2\gamma^*+\zeta^*,
\label{USF}	
\end{equation}
where $\eta^*\equiv \eta/\eta_0$, $a^*\equiv a/\nu$, $\gamma^*\equiv \gamma/(m \nu)$ and $\zeta^*\equiv \zeta/\nu$. Here, $\eta_0=p/\nu$ is the Navier-Stokes shear viscosity of a dilute (elastic) gas. Since $\eta^*$ and $\zeta^*$ are expected to be functions of the (reduced) shear rate $a^*$, the (reduced) friction coefficient $\gamma^*$ and the coefficient of restitution $\al$, Eq.\ \eqref{USF} establishes a relation between $a^*$, $\gamma^*$ and $\al$ and hence, only \emph{two} of them can be independent. Here, we will take $\gamma^*$ and $\al$ as the relevant (dimensionless) parameters measuring the departure of the system from equilibrium.

Before closing this Subsection, it is instructive to display the results derived for the granular suspension in the Navier-Stokes domain (small values of the shear rate). In this regime, the normal stress differences are zero and the form of the shear viscosity coefficient is \cite{GTSH12}
\beq
\label{2.19}
\eta_\text{NS}=\frac{nT}{\nu_\eta-\frac{1}{2}\left(\zeta_0-\frac{2}{m}\gamma\right)},
\eeq
where $\zeta_0$ is given by Eq.\ \eqref{2.18.1} and the collision frequency $\nu_\eta$ is \cite{BDKS98}
\begin{equation}
\label{2.21}
\nu_\eta=\frac{3\nu}{4d}\left(1-\alpha+\frac{2}{3}d\right)(1+\alpha).
\end{equation}
In Eqs.\ \eqref{2.18.1}, \eqref{2.19} and \eqref{2.21}, for the sake of simplicity, we have neglected non-gaussian corrections (proportional to the fourth cumulant) to $\zeta_0$, $\eta$ and $\nu_\eta$, respectively.

\subsection{Characteristic time scales and dimensionless numbers}
\label{time_scales}

\begin{figure}
\includegraphics[width=0.99 \columnwidth,angle=0]{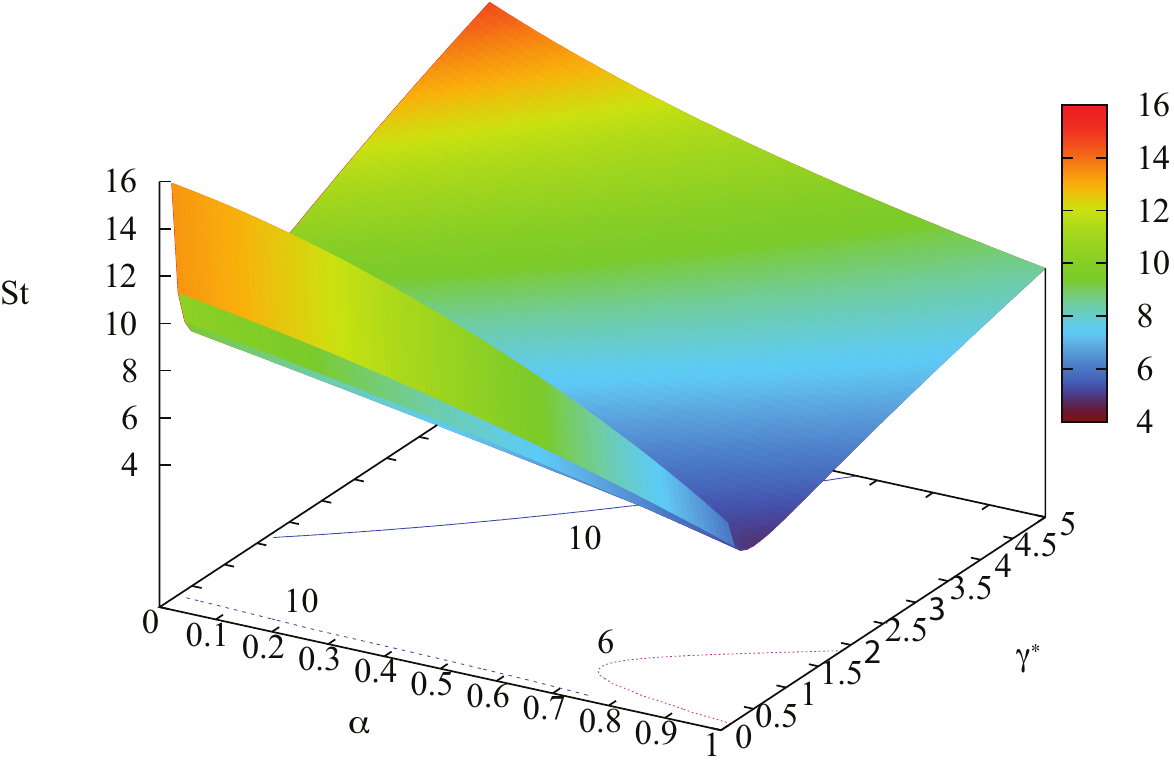}
\caption{$\mathrm{St}(\alpha,\gamma^*)$ surface for a dilute suspension of granular particles. The contours for $\mathrm{St}=6, 10$ have been marked in the $\mathrm{St}=0$ plane. \label{surface}}
\end{figure}
\begin{figure}
\includegraphics[width=0.90 \columnwidth,angle=0]{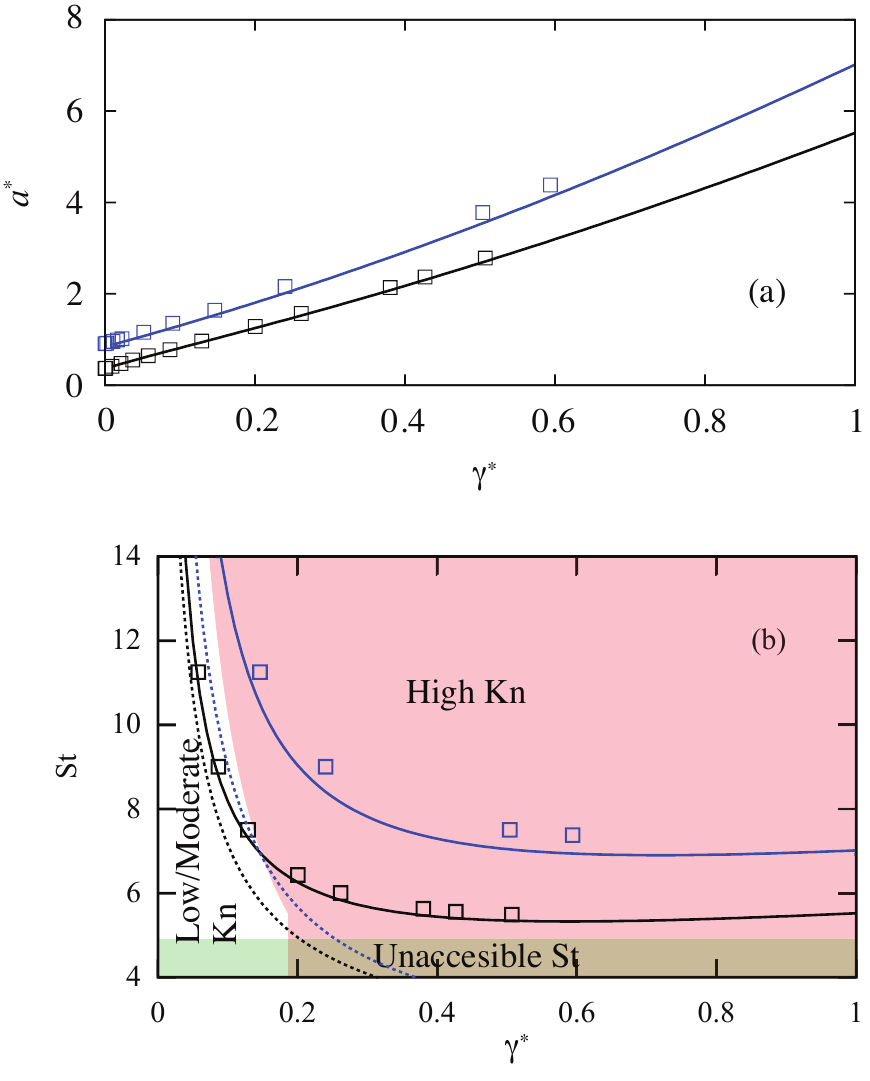}
\caption{Scheme of the flow regimes as they result from the relation \eqref{USF} between the (reduced) shear rate $a^*$, the (reduced) friction coefficient $\gamma^*$ and the Stokes number $\text{St}$ for a dilute granular suspension under USF. Blue (symbols and lines) stands for the case $\alpha=0.5$  and black (symbols and lines) stands for the case $\alpha=0.9$. The solid lines correspond to the results derived from Grad's moment method while the dashed lines refer to the Navier-Stokes predictions. Panel (a): Reduced shear rate $a^*$ vs. $\gamma^*$. Panel (b) Stokes number $\mathrm{St}$ vs. $\gamma^*$. In this panel the three regions commented in the text have been marked: a high Knudsen number region to the right of the panel (in pale red); a low/moderate Knudsen number region (in white) and finally, in the lower part of the panel, the forbidden small $\mathrm{St}$ region (green) may be found.  \label{class}}
\end{figure}

As it is known, in general there is more than one independent reduced length or time scale in a real flow problem (and thus, more than one independent Knudsen number \cite{B94}). Thus, let us analyze the dimensionless energy balance equation \eqref{USF}. It contains three \emph{homogeneous} terms, each one of them stands for the inverse of the three relevant (dimensionless) time scales of the USF problem, each caused with a different physical origin: the first term is proportional to the (reduced) shear rate $a^*$ that, according to its definition, is the shearing rate time scale (let us call it $\tau_s$); the second term is proportional to $\gamma^*$, thus setting the drag friction time scale ($\tau_d$); and finally, the third one, $\zeta^*$ comes from the inelastic cooling characteristic time scale ($\tau_i$).

A relevant dimensionless number in fluid suspensions is the Stokes number $\text{St}$ \cite{B74}. As in previous works \cite{TK95,SMTK96}, it is defined as the relation between the inertia of suspended particles and the viscous drag characteristic time scale :
\beq
\label{2.22}
\text{St}=\frac{m a}{3\pi \sigma \mu_g},
\eeq
where we recall that $\mu_g$ is the gas viscosity. According to Eq.\ \eqref{2.4}, $\text{St}$ can be easily expressed in terms of $\gamma^*$ and $a^*$ as
\begin{equation}
\label{Stokes}
\mathrm{St}=\frac{a^*}{\gamma^*/R_\text{diss}},
\end{equation}	
where $R_\text{diss}=1$ for dilute suspensions ($\phi=0$). Note that the Stokes number is a relevant parameter in fluid suspensions \cite{B74} since it measures the competition between the shearing and viscous friction mechanisms ($a^*$ and $\gamma^*$) on its rheological properties.

Since the reduced time scales ($\tau_s$, $\tau_d$, and $\tau_i$) have been defined with the inverse collision frequency $\nu^{-1}$, they may be regarded also as the characteristic Knudsen numbers ($\mathrm{Kn}$) of the system. For this reason, it is a necessary precondition for a Navier-Stokes hydrodynamic description of the problem (valid only for small enough spatial gradients), that all of them are small. In other words, as soon as one of them (just one) is close to one or higher, the Navier-Stokes approximation is expected to fail \cite{VSG13}.

However, as said before, only two of the relevant Knudsen numbers are actually independent since they are related through Eq.\ \eqref{USF}. For this reason, we additionally need to explore the relation between and $\tau_s$, $\tau_d$, and $\tau_i$ in order to analyze the limits of a Navier-Stokes description for the granular suspension under USF. For this, the reduced energy balance equation \eqref{USF} can be written in a perhaps more meaningful way for granular suspensions as a function of the Stokes number $\mathrm{St}$, namely,
\begin{equation}
-\frac{2}{d}\eta^*a^{*} +2\mathrm{St}^{-1}+\frac{\zeta^*}{a^*}=0.
\label{LTu2}
\end{equation}
Once the (scaled) non-Newtonian shear viscosity $\eta*$ and the (scaled) cooling rate $\zeta^*$ are given in terms of both $\alpha$ and $\gamma^*$, one can obtain the (scaled) shear rate $a^*$ (or equivalently, the reduced temperature $T^*\equiv \nu^2/a^2= a^{*-2}$) by solving the energy equation \eqref{LTu2}. This yields a cubic equation for $T^{*1/2}$ and has therefore three roots. A detailed study of the behavior of these roots has been previously made by Tsao and Koch \cite{TK95} for elastic suspensions and by Sangani \emph{et al.} \cite{SMTK96} for inelastic systems. The analysis shows that in general only one root is real at high values of the Stokes number while the other two are zero and negative (\emph{unphysical} solution). We focus now on the physical solution with positive temperature (that corresponds to the \textit{ignited} state of \cite{SMTK96}) by using the more general nonlinear Grad's solution derived in Sec.\ \ref{sec3}.

In Fig.\ \ref{surface} we plot the surface $\mathrm{St}(\alpha,\gamma^*)$  verifying Eq.\ \eqref{LTu2}. According to Fig.\ \ref{surface}, it is quite evident that it is not possible to reach a null value of the Stokes number. This is consistent with the energy equation \eqref{LTu2} since the latter value would imply $\text{St}^{-1}\to \infty$ and so, a balance between the different effects would not be possible. Figure \ref{class} (a) is the representation of two constant $\alpha$ curves of this surface, as obtained from the nonlinear Grad's solution (solid lines), explained in section \ref{sec3a}, and Monte Carlo simulations (symbols) for $d=3$ (spheres). The Navier-Stokes prediction for $\text{St}(\gamma^*)$ obtained from Eqs.\ \eqref{USF} and \eqref{2.19} is also plotted (dashed lines) for the sake of comparison. In Figure \ref{class} (b) we have marked with different colors three different regions: white stands for the region with $a^* \lesssim 1$, where the Navier-Stokes description is expected to apply (or in other words, where non-Newtonian corrections to rheological properties would not be significant), whereas red stands for the region where the Navier-Stokes approximation is expected to fail ($a^*\gtrsim 1$). The inelastic time scale $\tau_i$ would keep small as long as we do not represent too large inelasticity values. The drag time scale $\tau_d$ (or equivalently $\gamma^*$) is represented here only below 1. Thus, the only concern would be tracking small enough values of $\tau_s$ (or equivalently $a^*$) values. For this reason, the moderate/large $\mathrm{Kn}$ regions in Fig. \ref{class} (b) are separated by the curve that follows from the value $\gamma^*(\alpha,a^*=1)$ extracted from Eq. \eqref{LTu2}. Dark green region denotes the low $\mathrm{St}$ region that is not accessible for hydrodynamics (negative solutions for $T^{*1/2}$). As we can see in both panels (a) and (b), the agreement between Grad's solution (which takes into account non-Newtonian corrections to the shear viscosity) and simulations is excellent as long as keep in the small $\mathrm{Kn}$ region (both $\gamma^*<1$ and $a^*<1$).

The accuracy of Grad's solution extends deep inside the large $\mathrm{Kn}$ region, specially for lower inelasticities (note the black curve and symbols in the pale red region of Fig.\ \ref{surface}). On the other hand, as expected, the Navier-Stokes prediction exhibits significant discrepancies with simulations when $\text{Kn}\gg 1$. Please note that, although this is somewhat masked in the small range of values of $\gamma^*$ considered in Fig.\ \ref{class} (b), the Stokes number $\mathrm{St}$ is always a bi-valuated function of the (scaled) friction coefficient $\gamma^*$, as it can be clearly seen in Fig.\ \ref{surface}. Also notice from Fig.\ \ref{surface} that $\mathrm{St}$ has always a minimum with respect to $\gamma^*$ (at a given value of $\alpha$), although for scale reasons it is not very noticeable in Fig.\ \ref{class} (b).

It is important to finally remark in this section that the need for more complex constitutive equations (namely, those provided by Grad's moment method) is not a signal of a breakdown of hydrodynamics \cite{G03,TG98}, only a failure of the Navier-Stokes approximation \cite{DB99,VU09}. Also, let us note as an  important feature not described previously that $\mathrm{St}(\gamma^*)$ has two roots for each $\mathrm{St}$ value, as we can see in Fig. \ref{surface}.




\section{Theoretical approaches}
\label{sec3}

\subsection{Grad's moment method of the Boltzmann equation}
\label{sec3a}

We are interested here in obtaining the explicit forms of the relevant elements of the (scaled) pressure tensor $P_{ij}^*$ for a dilute granular suspension in terms of $a^*$, $\gamma^*$ and $\al$. To get it, we multiply both sides of Eq.\ \eqref{2.16} by $mV_iV_j$ and integrate over velocity. The result is
\beq
\label{3.13}
a_{ik}P_{kj}+a_{jk}P_{ki}+\frac{2\gamma}{m} P_{ij}=\Lambda_{ij}.
\eeq
where
\beq
\label{3.14}
\Lambda_{ij}\equiv \int\; \dd \mathbf{V}\; m V_iV_j J[\mathbf{V}|f,f],
\eeq
and we recall that $a_{ij}=a\delta_{ix}\delta_{jy}$. The exact expression of the collision integral $\Lambda_{ij}$ is not known, even in the elastic case. However, a good estimate can be expected by using Grad's approximation \cite{G49}
\begin{equation}
\label{3.1}
f(\mathbf{V})\to f_\text{M}(\mathbf{V}) \left(1 +\frac{m}{2nT^2}V_iV_j \Pi_{ij}\right),
\end{equation}
where
\begin{equation}
\label{3.11}
f_\text{M}(\mathbf{V})=n\left(\frac{m}{2\pi T}\right)^{d/2}e^{-mV^2/2T}
\end{equation}
is the (local) equilibrium distribution function  and
\begin{equation}
\label{3.12}
\Pi_{ij}=P_{ij}-p\delta_{ij}
\end{equation}
is the traceless part of the pressure tensor. Upon writing the distribution function \eqref{3.1} we have take into account that the heat flux is zero in the USF and we have also neglected the contribution of the fourth-degree velocity moment to $f$. This contribution has been recently considered \cite{G13} for the calculation of the Navier-Stokes transport coefficients of a granular fluid at moderate densities.

The collisional moment $\Lambda_{ij}$ can be determined when Eq.\ \eqref{3.1} is inserted into Eq.\ \eqref{3.14}. After some algebra (see Appendix \ref{appA} for details), we obtain the expression of $\Lambda_{ij}$ for inelastic hard spheres ($d=3$)
\beqa
\label{3.15}
\Lambda_{ij}&=&-p\nu (1+\al)\left[\frac{5}{12}(1-\al)\delta_{ij}+\frac{3-\al}{4}\right.\nonumber\\
&\times &  \left.\left(\Pi_{ij}^*+\frac{1}{14}\Pi_{ik}^*\Pi_{kj}^*
\right)-\frac{5+3\al}{672}\Pi_{k\ell}^*\Pi_{k\ell}^*\delta_{ij}\right],\nonumber\\
\eeqa
where $\Pi_{ij}^*\equiv \Pi_{ij}/p$. In the case of inelastic hard disks ($d=2$), the expression of $\Lambda_{ij}$ is
\beqa
\label{3.16}
\Lambda_{ij}&=&-p\nu \frac{1+\al}{2}\left[(1-\al)\delta_{ij}+\frac{7-3\al}{4}\Pi_{ij}^*\right.\nonumber\\
& & \left.+\frac{3}{64}(1-\al)
\Pi_{k\ell}^*\Pi_{k\ell}^*\delta_{ij}\right].
\eeqa
As we noted before, we evaluate $\Lambda_{ij}$ by retaining \emph{all} the quadratic terms in the tensor $\Pi_{ij}^*$. In particular, Eq.\ \eqref{3.15} reduces to the simpler expression obtained by Sangani \emph{et al.} \cite{SMTK96} for $d=3$ if we suppress  the quadratic terms in $\Pi_{ij}^*$. Also, if we particularize Eq.\ \eqref{3.15} for $\al=1$
\beq
\label{3.15.1}
\Lambda_{ij}=-p \nu \left[\Pi_{ij}^*+\frac{1}{14}\left( \Pi_{ik}^*\Pi_{kj}^*-\frac{1}{3}\Pi_{k\ell}^*\Pi_{k\ell}^*\delta_{ij}\right)\right],
\eeq
and hence we recuperate the expression of $\Lambda_{ij}$ derived for Tsao and Koch \cite{TK95} for the special case of perfectly elastic particles (see Eq.\ (3.7) of \cite{TK95}). Thus, our expression \eqref{3.15} for the collisional moment $\Lambda_{ij}$ for inelastic hard spheres is more general and can recover the results of previous bibliography.

In addition, we have also checked that the expression \eqref{3.15} agrees with a previous and independent derivation of $\Lambda_{ij}$ for inelastic hard spheres \cite{GT12}. This shows the consistency of our nonlinear Grad's solution.

The nonlinear contribution $\zeta_2$ to the cooling rate [defined by Eq.\ \eqref{cooling_int}] can be obtained for spheres and disks from Eqs.\ \eqref{3.15} and \eqref{3.16}, respectively:
\beq
\label{3.17}
\zeta_\text{spheres}^*=\frac{5}{12}(1-\al^2)\left(1+\frac{1}{40}\Pi_{k\ell}^*\Pi_{k\ell}^*\right),
\eeq
\beq
\label{3.18}
\zeta_\text{disks}^*=\frac{(1-\al^2)}{2}\left(1+\frac{3}{64}\Pi_{k\ell}^*\Pi_{k\ell}^*\right).
\eeq
Here again, this is a more general and accurate expression of the cooling rate for dilute granular suspensions. Of course, for elastic collisions ($\al=1$), we recover the limit $\zeta^*=0$ \cite{TK95}. Moreover in the linear in $\Pi_{ij}^*$ approach, $\zeta^*\to (5/12)(1-\al^2)$ for spheres, which agrees with the previous results \cite{SMTK96}.

The knowledge of the collisional moment $\Lambda_{ij}$ allows us to get the explicit form of the relevant elements of the pressure tensor $P_{ij}^*$. Their forms are provided in the Appendix \ref{appA}.


\subsection{BGK-type kinetic model of the Boltzmann equation}
\label{sec3b}

Now we consider the results derived for the USF from a BGK-type kinetic model of the Boltzmann equation \cite{BDS99}. In the USF problem, the steady kinetic model for the granular suspension described by the Boltzmann equation \eqref{2.16} becomes
\beq
\label{3.26}
-aV_y\frac{\partial f}{\partial V_x}-\frac{\gamma}{m} \frac{\partial}{\partial
{\bf V}}\cdot {\bf V} f =-\chi(\alpha)\nu \left(f-f_\text{M}\right)+\frac{\zeta_0}{2}\frac{\partial}{\partial
{\bf V}}\cdot {\bf V} f,
\eeq
where $\nu$ is the effective collision frequency defined by Eq.\ \eqref{nu}, $f_\text{M}$ is given by Eq.\ \eqref{3.11}, $\zeta_0$ is defined by Eq.\ \eqref{2.18.1}, and $\chi(\alpha)$ is a free parameter of the model chosen to optimize the agreement with the Boltzmann results.

One of the main advantages of using the kinetic model \eqref{3.26} instead of the Boltzmann equation is that it lends itself to get an exact solution. The knowledge of the form of $f(\mathbf{V})$ allows us to determine \emph{all} its velocity moments. The explicit forms of the distribution function $f(\mathbf{V})$ as well as its moments are provided in the Appendix \ref{appB}. In particular, the relevant elements of the pressure tensor are given by
\beq
\label{3.39}
\Pi_{yy}^*=\Pi_{zz}^*=-\frac{2\widetilde{\epsilon}}{1+2\widetilde{\epsilon}},
\quad \Pi_{xy}^*=-\frac{\widetilde{a}}{(1+2\widetilde{\epsilon})^2},
\eeq
where the (dimensionless) shear rate $\widetilde{a}$ obeys the equation
\beq
\label{3.40}
\widetilde{a}^2=d\widetilde{\epsilon}(1+2\widetilde{\epsilon})^2.
\eeq
Here, $\widetilde{a}\equiv a^*/\chi$, $\widetilde{\zeta}\equiv \zeta^*/\chi$, $\widetilde{\epsilon}\equiv \widetilde{\gamma}+\widetilde{\zeta}/2$, and  $\widetilde{\gamma}\equiv \gamma^*/\chi$. The expressions \eqref{3.39} and \eqref{3.40} are fully equivalent to linear Grad's predictions \eqref{3.23}-\eqref{3.24.1}, except that $\chi$ is replaced by $\beta$.

\section{Numerical solutions: Direct simulation Monte Carlo method}
\label{sec4}

As we said in the Introduction, the third method consists in obtaining a numerical solution to the Boltzmann equation \eqref{2.16} by means of the DSMC method \cite{B94} applied to inelastic hard spheres. More concretely, the algorithm we used is analogous to the one employed in Ref.\ \cite{MGSB99} where the USF state becomes homogeneous in the frame moving with the flow velocity $\mathbf{U}$. Here, we have simply added the drag force coming from the interaction between the solid particles and the surrounding interstitial fluid. The initial state is the same for all simulations, namely, Gaussian velocity distributions with homogeneous density and temperature. We have observed in most of the cases that, after a relatively short transient, a steady state is reached. In this state, the relevant quantities of the USF problem (nonzero elements of the pressure tensor, the kurtosis and the velocity distribution function) are measured.

Since the base of the algorithm has been explained in detail in previous papers \cite{MGSB99,CVG13}, we skip here these details and only comment that we have performed systematic simulation series for two different situations: (i) by varying the (scaled) friction coefficient $\gamma^*$ at a given value of $\alpha$ and, conversely, (ii) by varying the coefficient of restitution $\alpha$ at a given value of $\gamma^*$. In addition, the series corresponding to varying $\gamma^*$ have been employed for graphs with varying the Stokes number $\mathrm{St}$ also.

The use of the DSMC method is convenient since it is considered as an accurate method of solving the Boltzmann equation. Here, the DSMC results can be considered as a clean way to assess the degree of reliability of the theoretical descriptions we developed (Grad's moment method and BGK-type kinetic model). This is what we do, along with presentation of the results, in the following Section.

\section{Results}
\label{sec5}

We devote this Section to direct comparative presentation of the results obtained from all three independent routes we have followed for this work. Although our theoretical expressions apply for spheres and disks, for the sake of brevity we present only results for the physical case of a three-dimensional system ($d=3$). Given that the computational algorithm can be easily adapted to disks, a comparison between theory and simulation for $d=2$ could be also performed.

\begin{figure}
\includegraphics[width=0.90 \columnwidth,angle=0]{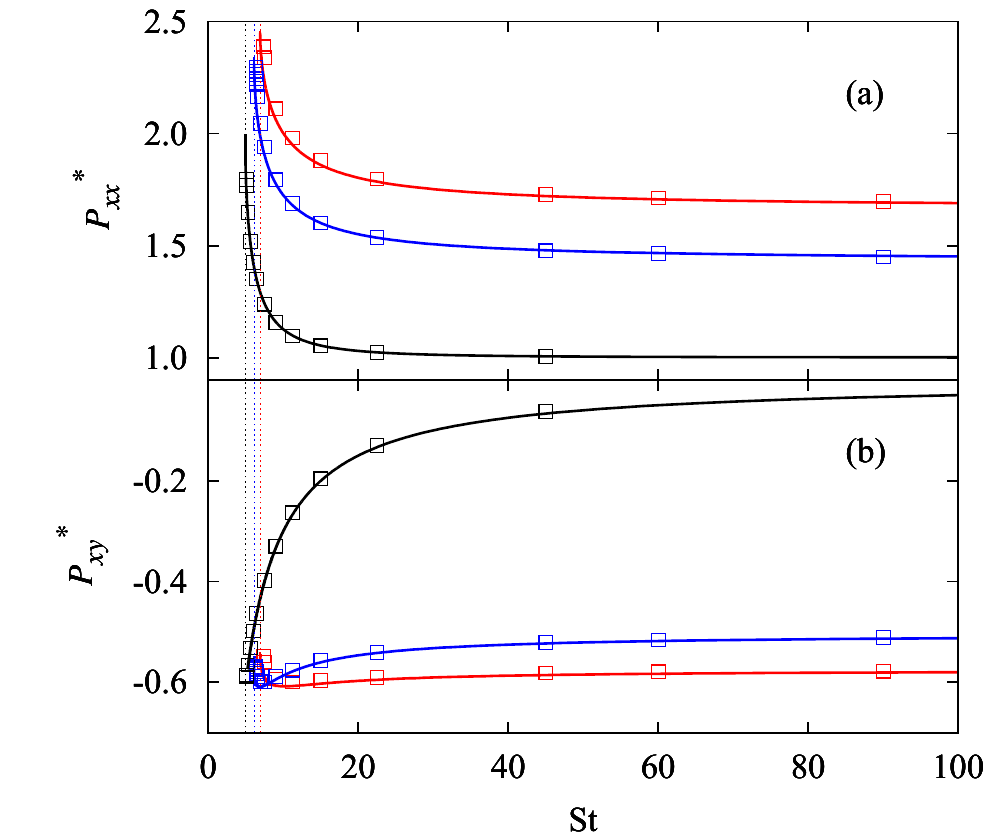}
\caption{Dependence of the (reduced) elements of the pressure tensor $P_{xx}^*$ (panel (a)) and $P_{xy}^*$ (panel (b)) on the Stokes number $\text{St}$ for several values of the coefficient of restitution $\al$: $\alpha=1$ (black), $\alpha=0.7$ (blue) and $\alpha=0.5$ (red). The solid lines are the theoretical results obtained from nonlinear Grad's solution while the symbols refer to the results obtained from Monte Carlo simulations. We have marked as vertical dotted lines the minimum allowed value for the Stokes number $\mathrm{St}$.\label{figPij}}
\end{figure}
\begin{figure}
\includegraphics[width=0.90 \columnwidth,angle=0]{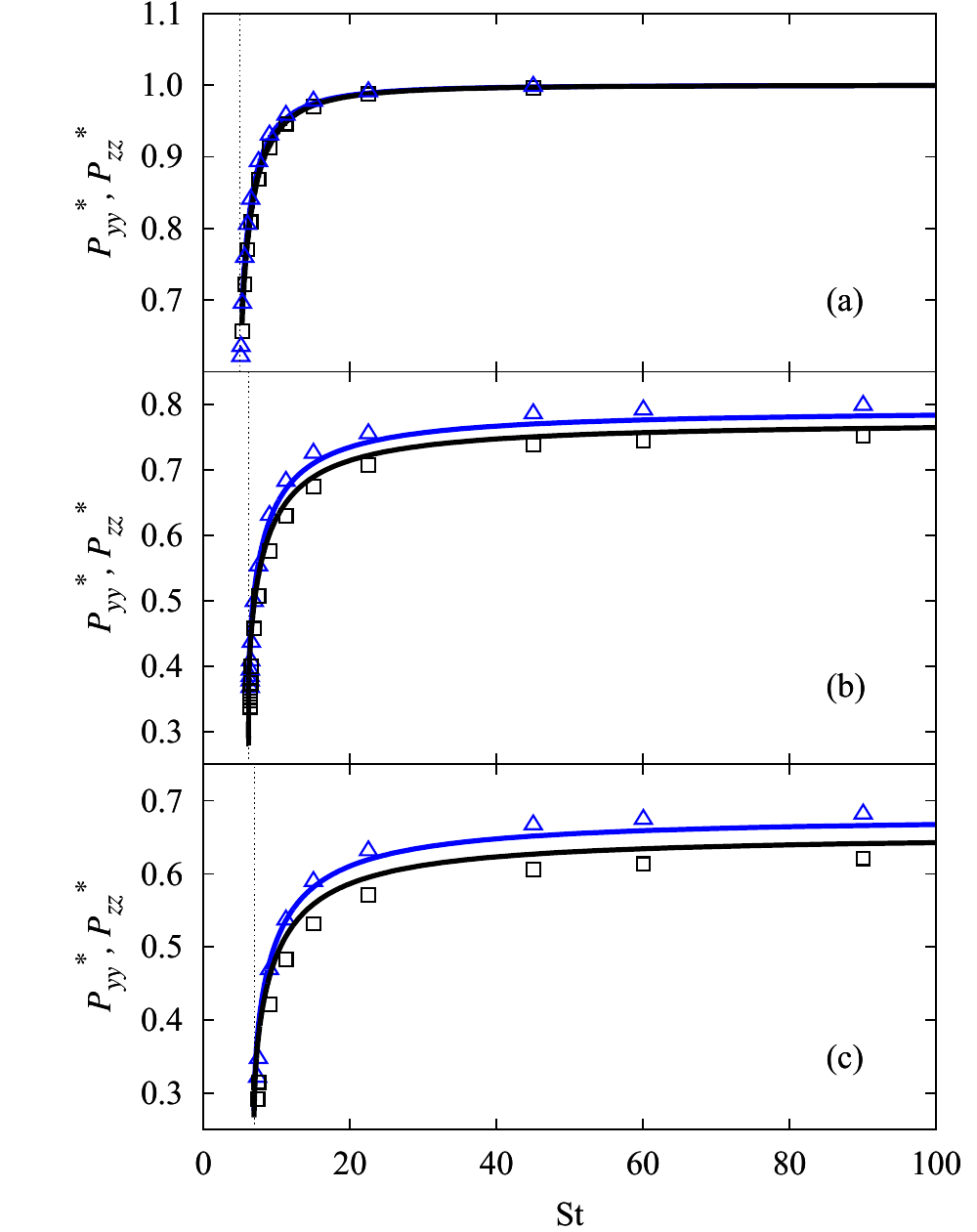}
\caption{Dependence of the (reduced) diagonal elements of the pressure tensor $P_{yy}^*$ (black lines and squares) and $P_{zz}^*$ (blue lines and triangles) on the Stokes number $\text{St}$ for several values of the coefficient of restitution $\al$: $\alpha=1$ (a), $\alpha=0.7$ (b) and $\alpha=0.5$ (c). The solid lines are the theoretical results obtained from nonlinear Grad's solution while the symbols refer to the results obtained from Monte Carlo simulations. As in Fig.\ \ref{figPij}, we have marked as vertical dotted lines the minimum allowed value of the Stokes number $\mathrm{St}$ for each value of $\al$.
\label{figPii}}
\end{figure}
\begin{figure}
\includegraphics[width=0.90 \columnwidth,angle=0]{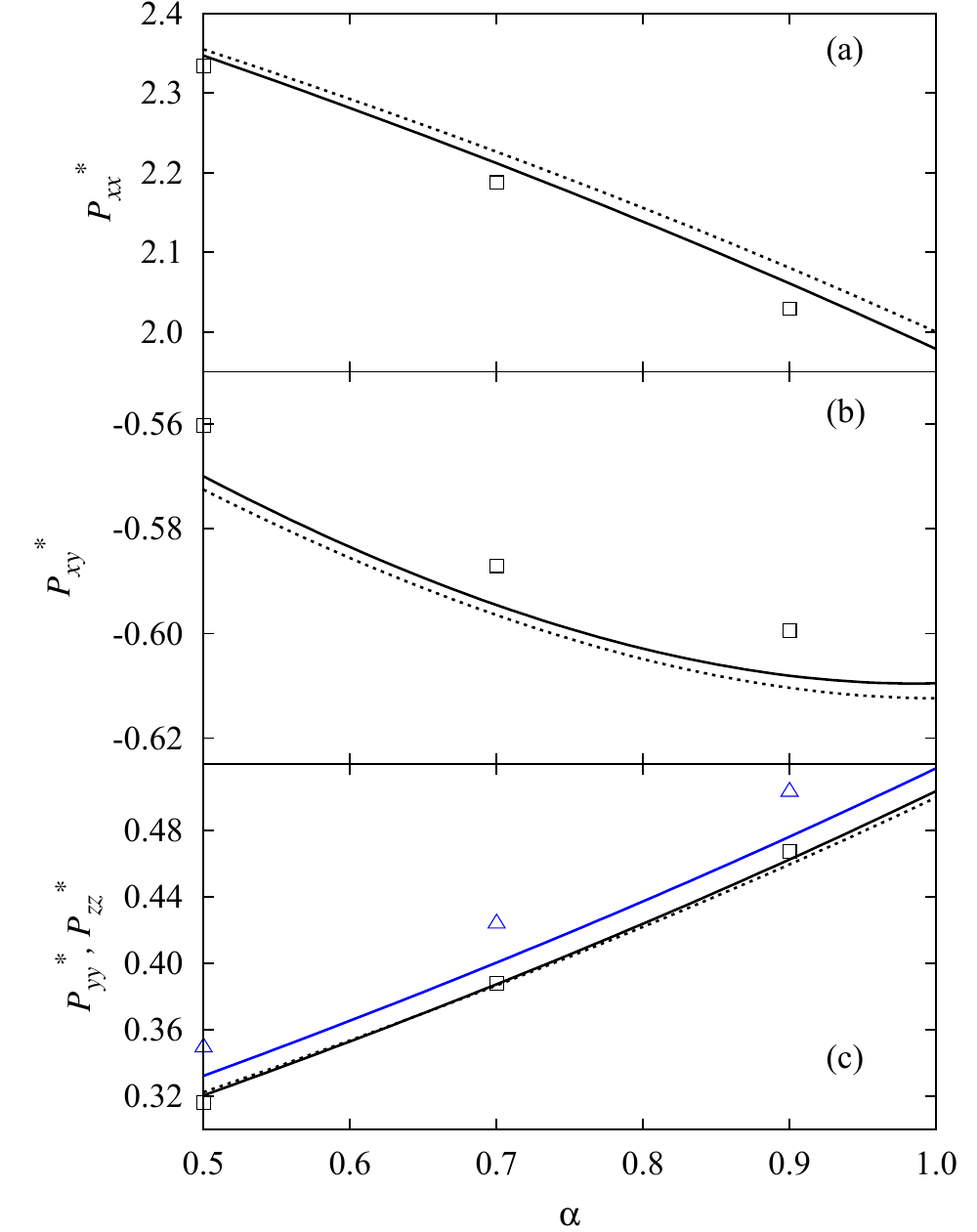}
\caption{Plot of the (reduced) nonzero elements of the pressure tensor $P_{xx}^*$ (panel a), $P_{xy}^*$ (panel b), $P_{yy}^*$ and $P_{zz}^*$ (panel c) as functions of the coefficient of restitution $\al$ for  $\gamma^*=0.5$. The solid and dotted lines correspond to the results obtained from nonlinear and linear Grad's solution, respectively. Symbols refer to Monte Carlo simulations. In the panel (c), the blue solid line and triangles are for the element $P_{zz}^*$ while the black solid line and squares are for the element $P_{yy}^*$. Note that linear Grad's solution (dotted line) yields $P_{yy}^*=P_{zz}^*$.
\label{figPijalfa}}
\end{figure}


Figure \ref{figPij} shows the dependence of the (reduced) elements $P_{xx}^*$ and $P_{xy}^*$ of the pressure tensor on the
Stokes number $\text{St}$. Here, we have performed simulation series by varying the (reduced) friction coefficient
$\gamma^*$ (or equivalently, $\text{St}$)  for three different values of the coefficient of restitution:
$\alpha=1$ (elastic case), $\al=0.9$ and $\al=0.7$. Recall that the diagonal elements of the pressure tensor
are related as $P^*_{xx}+P^*_{yy}+(d-2)P^*_{zz}=d$. In this graph, only the predictions given by the
so-called nonlinear Grad's solution are plotted. The results obtained from linear Grad's solution are practically indistinguishable from the latter ones for the cases considered in this plot. The comparison between theory (solid lines) and computer simulations (symbols) shows an excellent agreement for all values of the Stokes number represented here, independently of the degree of inelasticity of collisions in the granular gas.

As noted in the Introduction, one of the drawbacks of linear's Grad solution  is that yields $P_{yy}^*=P_{zz}^*$ and hence, the second viscometric function (proportional to $P_{yy}^*-P_{zz}^*$ \cite{BAH87}) vanishes. This failure of linear Grad's solution is also present at moderate densities (see Eq.\ (4.33) of \cite{SMTK96}).  Figure \ref{figPii} shows the dependence of the normal elements $P_{yy}^*$ and $P_{zz}^*$ on the Stokes number $\text{St}$ as obtained from the DSMC method (symbols) and nonlinear Grad's solution. It is quite apparent that both simulations and theory show that $P_{zz}^*>P_{yy}^*$. This is specially relevant in granular suspensions since we have two different sink terms ($\gamma^*$ and $\zeta^*$) in the energy balance equation \eqref{USF}. And thus, the non-Newtonian effects like $P_{yy}^*\neq P_{zz}^*$ are expected to be stronger. The balance of these two terms with the viscous heating term ($\eta^* a^{*2}$) requires high shear rates as can be seen in Fig.\ \ref{class}. We observe in Fig.\ \ref{figPii} that our theory captures quantitatively well the tendency of $P^*_{yy}$ (the diagonal element of the pressure tensor in the direction of shear flow) to become smaller than $P^*_{zz}$, this tendency being stronger as inelasticity increases (and disappearing completely in the elastic limit $\alpha=1$). It is also apparent that the dependence of both $P_{zz}^*$ and $P_{yy}^*$ on the Stokes number is qualitatively well captured by nonlinear Grad's solution, even for strong collisional dissipation. Finally, regarding rheology and as a complement of Figs.\ \ref{figPij} and \ref{figPii}, Fig. \ \ref{figPijalfa} shows the $\al$-dependence of the relevant elements of the pressure tensor at a given value of the (scaled) friction coefficient $\gamma^*$. Since the value of $\gamma^*$ is relatively high ($\gamma^*=0.5$), the results presented in Fig.\ \ref{figPijalfa} can be considered as an stringent test for both linear and nonlinear Grad's solutions. Although linear Grad's solution exhibits a reasonably good agreement with DSMC data, we see that nonlinear Grad's solution mitigates in part the discrepancies observed by using the linear approach since the former theory correctly predicts the trend of the normal stress difference $P_{zz}^*-P_{yy}^*$ and also improves the agreement with simulations for the elements $P_{xx}^*$ and $P_{xy}^*$. On the other hand, since the system is quite far from equilibrium, there are still quantitative discrepancies between the nonlinear theory and simulations.

\begin{figure}
\includegraphics[width=0.90 \columnwidth,angle=0]{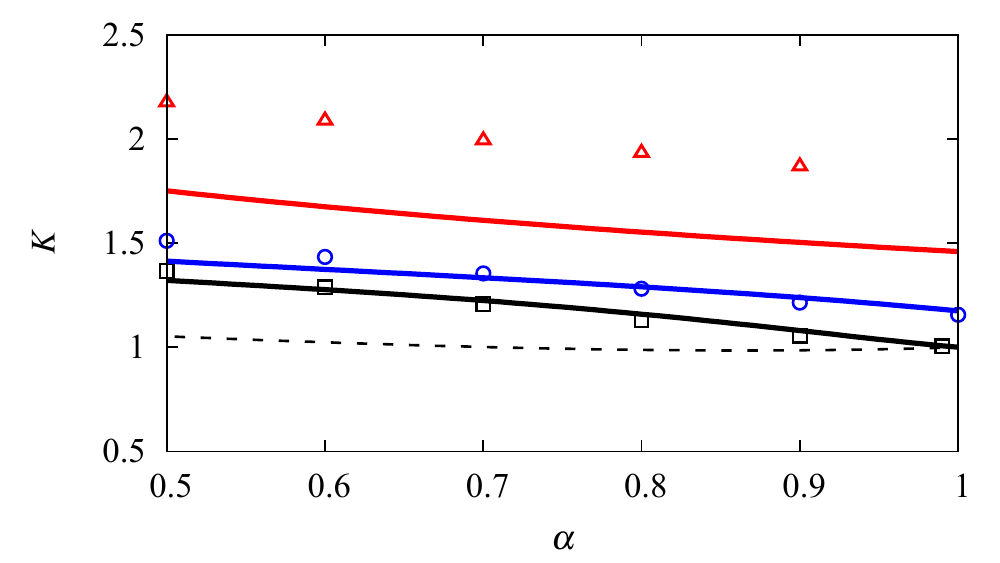}
\caption{Plot of the kurtosis $K\equiv \langle V^4 \rangle/\langle V^4 \rangle_0$ versus the coefficient of restitution $\al$ for three different values of the (reduced) friction coefficient $\gamma^*$: $\gamma^*=0$ (black line and squares), $\gamma^*=0.1$ (blue line and circles) and $\gamma^*=0.5$ (red line and triangles). The solid lines correspond to the results obtained from the BGK-type model while symbols refer to DSMC results. The dashed line is the result obtained in Ref.\ \cite{GTSH12} for the homogeneous cooling state \label{kurtosis}.}
\end{figure}

Next, we present results for the kurtosis or fourth order cumulant $K\equiv \langle V^4 \rangle/\langle V^4 \rangle_0$ where
\beq
\label{K}
\langle V^k \rangle=\frac{1}{n}\int\; \dd \mathbf{V} V^k f(\mathbf{V}),
\eeq
and
\beq
\label{K0}
\langle V^k \rangle_0=\frac{1}{n}\int\; \dd \mathbf{V} V^k f_\text{M}(\mathbf{V}).
\eeq
The dependence of the kurtosis on both $\gamma^*$ and $\al$ can be easily obtained from the results derived from the BGK-type kinetic model [see Eq.\ \eqref{3.38} for the BGK velocity moments].
Note that $\langle V^k \rangle=\langle V^k \rangle_0$ if one uses Grad's distribution \eqref{3.1}, which is a failure of Grad's solution since $K$ is clearly different from 1. Figure \ref{kurtosis} shows the dependence of $K$ on the coefficient of restitution $\al$ for hard spheres ($d=3$) and three different values of the (reduced) friction coefficient $\gamma^*$: $\gamma^*=0$ (dry granular gas),  $\gamma^*=0.1$ and $\gamma^*=0.5$. In the case of elastic collisions ($\al=1$), $K=1$ only for $\gamma^*=0$ since in this case the system is at equilibrium ($f=f_\text{M}$). We have also included the result obtained in Ref.\ \cite{GTSH12} in the HCS, which is independent of $\gamma^*$. It is important to remark first that the simulation results obtained independently here for $\gamma^*=0$ in Fig.\ \ref{kurtosis} are consistent with those previously reported for a sheared granular gas with no interstitial fluid \cite{AS05}. For low values of $\gamma^*$, we see that the agreement between theory and simulation is very good in the full range of values of inelasticities represented here. This shows again the reliability of the BGK model to capture the main trends observed in granular suspensions. On the other hand, the agreement is only qualitative for relatively high values of the friction coefficient $\gamma^*$ since the BGK results clearly underestimate the value of the kurtosis given by computer simulations. These discrepancies between the BGK-type model and DSMC for the fourth-degree velocity moment in non-Newtonian states is not surprising since the above kinetic model does not intend to mimic the behavior of the \emph{true} distribution function beyond the thermal velocity region. As expected, it is apparent that the prediction for $K$ in the homogeneous state differs clearly from the one obtained in the DSMC simulations at $\gamma^*=0$.
\begin{figure}
\includegraphics[width=0.9 \columnwidth,angle=0]{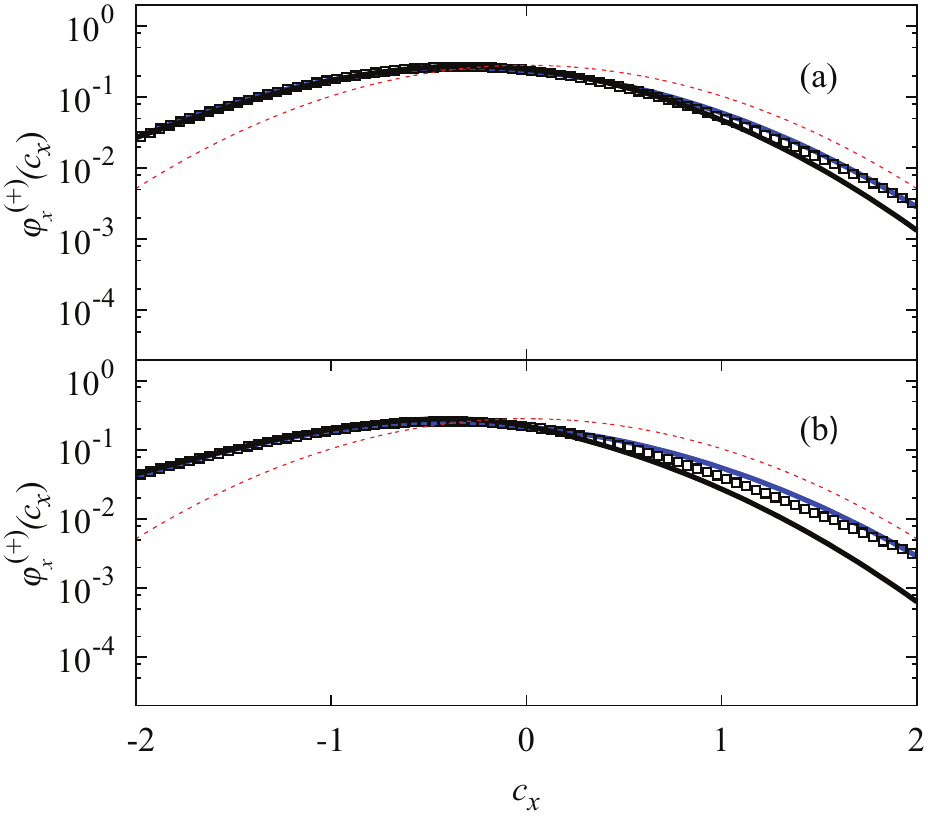}
\caption{(Color online) Logarithmic plots of the marginal distribution function $\varphi_x^{(+)}(c_x)$, as defined in Eq.\ \eqref{phix}. Two cases are represented here: (a) $\alpha=0.9$, $\gamma^*=0.1$ and (b) $\alpha=0.5$, $\gamma^*=0.1$. The black and blue solid lines are the theoretical results derived from the BGK model and the ME formalism, respectively, while the symbols represent the simulation results. The red dotted lines are the (local) equilibrium distributions.    \label{fxre}}
\end{figure}
\begin{figure}
\includegraphics[width=0.90 \columnwidth,angle=0]{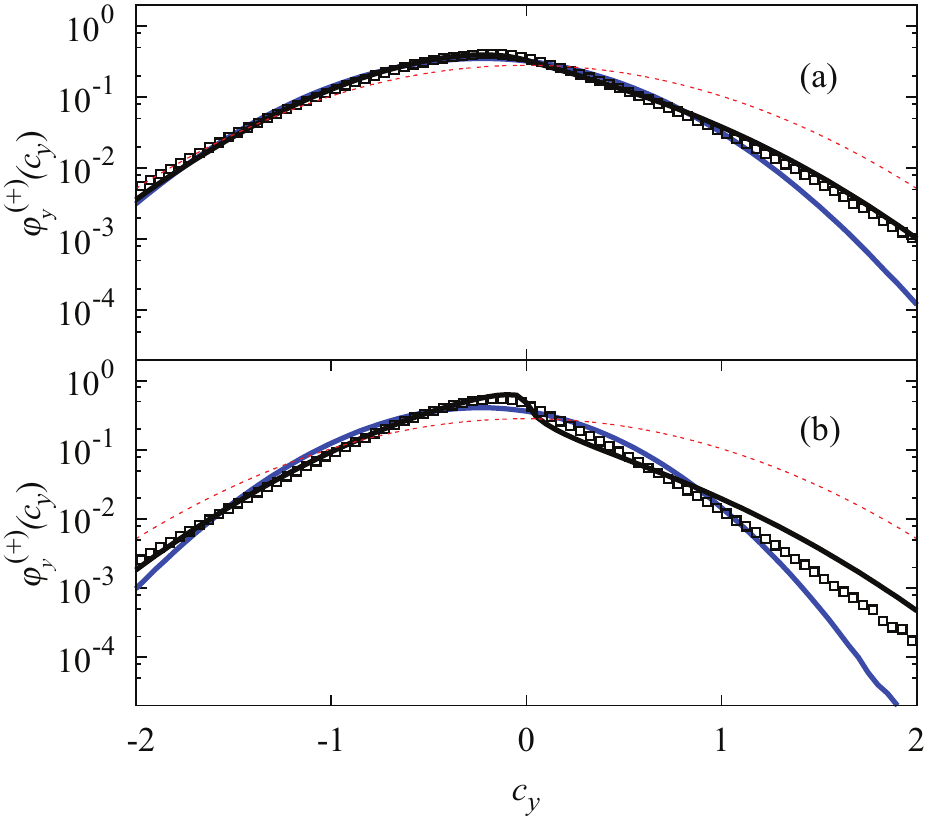}
\caption{(Color online) Logarithmic plots of the marginal distribution function $\varphi_y^{(+)}(c_y)$, as defined in Eq.\ \eqref{phiy}. Two cases are represented here: (a) $\alpha=0.9$, $\gamma^*=0.1$ and (b) $\alpha=0.5$, $\gamma^*=0.1$. The black and blue solid lines are the theoretical results derived from the BGK model and the ME formalism, respectively, while the symbols represent the simulation results. The red dotted lines are the (local) equilibrium distributions. \label{fyre}}
\end{figure}

Apart from the rheological properties and the high velocity moments, the solution to the BGK-type model provides the explicit form of the velocity distribution function $f(\mathbf{V})$. Figures \ref{fxre} and \ref{fyre} show the marginal distributions $\varphi_x^{(+)}(c_x)$ [defined by Eq.\ \eqref{phix}] and $\varphi_y^{(+)}(c_y)$ [defined by Eq.\ \eqref{phiy}], respectively, for $\gamma^*=0.1$ and two different values of the coefficient of restitution $\al$: $\al=0.9$ (moderate inelasticity) and $\al=0.5$ (strong inelasticity). The black solid lines are the results derived from the BGK model and the symbols represent Monte Carlo simulations. For the sake of completeness, it is interesting to use the maximum-entropy (ME) formalism \cite{BM91} to construct the distribution maximizing the functional
\beq
\label{ME}
-\int\; \dd \mathbf{V}\; f(\mathbf{V})\; \ln f(\mathbf{V}),
\eeq
subjected to the constraints of reproducing the density $n$ and the pressure tensor $\mathsf{P}$. In the three-dimensional case, this yields
\beq
\label{ME.1}
f(\mathbf{V})=n \pi^{-3/2} \det\left( \mathbf{Q} \right)^{1/2} \exp\left(-\mathbf{V}\cdot \mathbf{Q} \cdot \mathbf{V} \right),
\eeq
where $\mathbf{Q}\equiv \frac{1}{2}m n \mathsf{P}^{-1}$. The ME approximation \cite{GT78} was employed by Jenkins and Richman \cite{JR88} in order to determine the kinetic contributions to the pressure tensor in a sheared granular fluid of hard disks. Moreover, in Figs.\ \ref{fxre} and \ref{fyre}, as a reference the (local) equilibrium distributions (red dotted lines) are also represented. Although not shown in Figs.\ \ref{fxre} and \ref{fyre}, Grad's distribution \eqref{3.1} could lead to unphysical (negative) values of the marginal
distributions $\varphi_x^{(+)}(c_x)$ and $\varphi_y^{(+)}(c_y)$ for large velocities. This is again a drawback of Grad's solution not shared by the BGK solution since the latter is always positive definite for any range of velocities considered. Regarding the comparison between the different results, since the (reduced) shear rate is not small [see for instance, Fig.\ \ref{class} for $\al=0.5$ and $\gamma^*=0.1$], we observe that the distortion from the Gaussian distribution is quite apparent in the three different approaches (BGK, ME and DSMC). Two anisotropic features of the USF state are seen. First, the functions $\varphi_x^{(+)}(c_x)$ and $\varphi_y^{(+)}(c_y)$ are asymmetric since $\varphi_x^{(+)}(|c_x|)<\varphi_x^{(+)}(-|c_x|)$ and $\varphi_y^{(+)}(|c_y|)<\varphi_y^{(+)}(-|c_y|)$. This is a physical effect induced by the shearing since the shear stress $P_{xy}^*<0$. The second feature is the non-Newtonian property $\varphi_x^{(+)}(c_x)<\varphi_y^{(+)}(c_y)$. In fact, the marginal distribution $\varphi_x^{(+)}(c_x)$ is thicker than $\varphi_y^{(+)}(c_y)$, in consistency with the result $P_{xx}^*-P_{yy}^*>0$. The above two effects are more pronounced for $\al=0.5$ than for $\al=0.9$. With respect to the comparison between theory and simulation, we observe that in general the agreement between theoretical predictions (the BGK model and the ME formalism) and simulations data is excellent in the region of thermal velocities ($|c_i|\sim 1$). It is also apparent that while the ME approach compares better with simulations than the BGK results for the distribution $\varphi_x^{(+)}(c_x)$, the opposite happens for the distribution $\varphi_y^{(+)}(c_y)$. In particular, in the case of $\al=0.9$ the BGK model (the ME formalism) yields an excellent agreement with DSMC over the complete range of velocities studied for the distribution $\varphi_y^{(+)}(c_y)$ ($\varphi_x^{(+)}(c_x)$). On the other hand, for larger velocities and strong collisional dissipation, there are quantitative discrepancies between theoretical predictions and simulations.

\begin{figure}
\includegraphics[width=0.85 \columnwidth,angle=0]{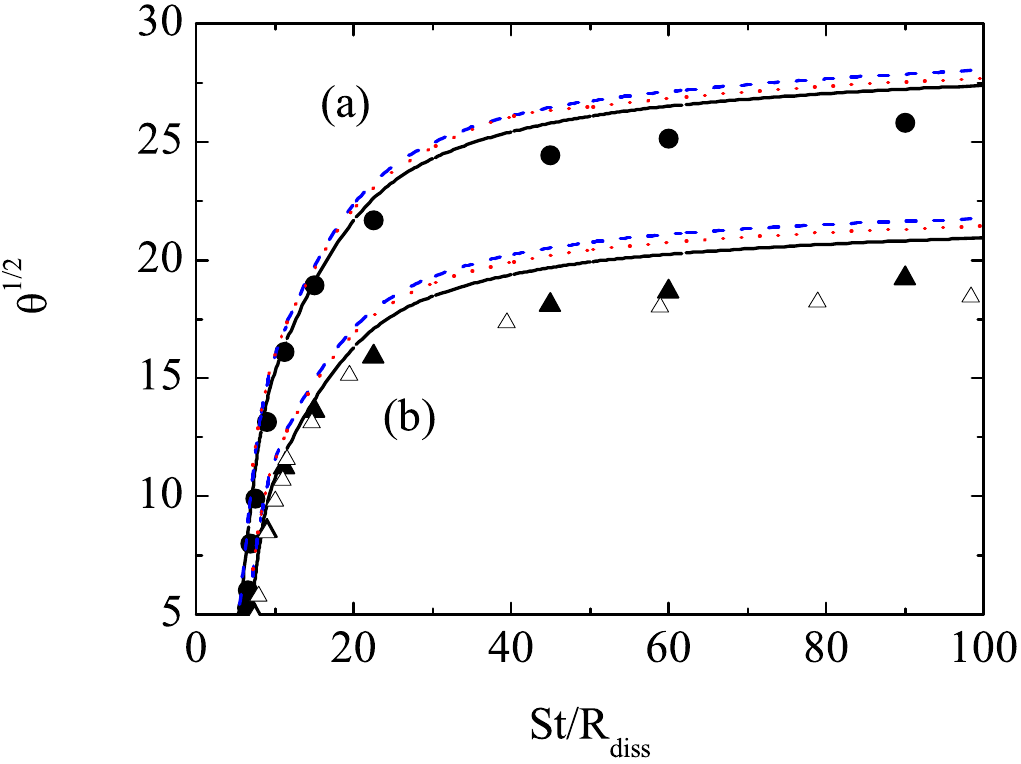}
\caption{Plot of the square root of the steady granular temperature $\theta^{1/2}$ as a function of $\text{St}/R_\text{diss}$ in the case of hard spheres ($d=3$) for $\phi=0.01$. Two different values of the coefficient of restitution have been considered: $\al=0.7$ (a) $\al=0.5$ (b). The solid line is the Grad solution (including nonlinear contributions) to the Boltzmann equation, the dashed (blue) line corresponds to the BGK results (which coincide with those obtained from the linear Grad solution) and the dotted (red) line refers to the results obtained by Sangani \emph{et al.} \cite{SMTK96} from the Enskog equation by applying (linear) Grad's method. The black circles and triangles are the simulation results obtained here by means of the DSMC method for $\al=0.7$ and $\al=0.5$, respectively, while the empty triangles are the results obtained in Ref.\ \cite{SMTK96}.  \label{fig5}}
\end{figure}

Finally, it is quite interesting to compare the dynamic simulation results reported in Ref.\ \cite{SMTK96} in the case of very dilute suspensions ($\phi=0.01$) with those carried here by means of the DSMC method. To do it, we introduce the (steady) granular temperature $\theta$ as
\beq
\label{4.4}
\theta=\frac{4T}{m\sigma^2 a^2}=\frac{25\pi}{2304}\frac{1}{\phi^2 a^{*2}},
\eeq
where we recall that $\phi=(\pi/6)n\sigma^3$ is the volume fraction for spheres. Figure \ref{fig5} shows $\sqrt{\theta}$ versus $\text{St}/R_\text{diss}$ for two different values of the coefficient of restitution: $\al=0.7$ and $\al=0.5$.  We have considered the Monte Carlo simulations performed here for $\al=0.7$ and $\al=0.5$ and those made in Ref.\ \cite{SMTK96} in the case $\al=0.5$. In addition, we have also included the theoretical results derived in \cite{SMTK96} from the Enskog equation.  We observe first that the dynamic simulations for finite Stokes number and the DSMC results are consistent among themselves in the range of values of $\text{St}/R_\text{diss}$ explored. This good agreement gives support to the applicability of the model for dilute granular suspensions introduced in Eq.\ \eqref{2.16}. It is also apparent that the performance of nonlinear Grad's theory for the (steady) temperature is slightly better than the remaining theories. Notice also that the agreement between theory and computer simulations improves as we approach to the dry granular limit $\mathrm{St}/\mathrm{R_{diss}}\to \infty$. Thus, at $\alpha=0.7$, for instance the discrepancies between nonlinear Grad's theory and DSMC results for $\mathrm{St}/\mathrm{R_{diss}}=11.3$, 22.5, 45, 60 and 90 are about of 8.5\%, 6.4\%, 5.8\%, 5.5\% and 5.4\%, respectively while at $\alpha=0.5$ the discrepancies are about of 14\%, 10\%, 9\%, 8.6\% and 8.5\%, respectively. This shows again that our Grad's solution compares quite well with simulations for not too large values of the (scaled) friction coefficient $\gamma^*$ (or equivalently, for large values of the Stokes number $\mathrm{St}$).

\section{Conclusions}
\label{sec6}

In this work, we have presented a complete and comprehensive theoretical description of the non-Newtonian transport properties of a dilute granular suspension under USF in the framework of the (inelastic) Boltzmann equation. The influence of the interstitial fluid on the dynamic properties of grains has been modeled via a viscous drag force proportional to the particle velocity. This type of external force has been recently employed in different works on gas-solid flows \cite{WZLH09,H13,HT13,SMMD13,WGZS14}.  Our study has been both theoretical and computational. In the theory part, we have presented results from two different approaches: Grad's moment method and a BGK-type kinetic model used previously in other granular flow problems and now applied specifically to the model of granular suspensions. In contrast to previous works in granular sheared suspensions \cite{SMTK96}, we have included in Grad's solution quadratic terms in the pressure tensor $P_{ij}$ in the collisional moment $\Lambda_{ij}$ associated with the momentum transport (nonlinear Grad's solution). This allows us to evaluate the normal stress differences in the plane normal to the laminar flow (namely, the normal stress difference $P_{yy}^*-P_{zz}^*$) and of course, one obtains more accurate expressions of the non-Newtonian transport properties. The inclusion of quadratic terms in $P_{ij}$ in the evaluation of $\Lambda_{ij}$ was already considered by Tsao and Koch \cite{TK95} in an analogous system but only in the limit of perfectly elastic collisions ($\al=1$). Therefore, for strictly granular particles (i.e., beyond the elastic limit) this is the first time that, to the best of our knowledge, the difference $P_{yy}^*-P_{zz}^*$ has been analytically detected and evaluated in a theory of sheared \textit{granular} suspensions. This is one of the most relevant achievements of the present contribution. Moreover, the development of the corresponding BGK-type model for the dilute granular suspension under shear has allowed us also to formally compute all velocity moments as well as the velocity distribution function of the suspension.


Additionally, to gauge the accuracy of the above theoretical approaches, we have presented simulation results (DSMC method applied to the inelastic Boltzmann equation). The comparison between theory and Monte Carlo simulations has been done by varying both the (scaled) friction coefficient $\gamma^*$ (or equivalently, the Stokes number $\text{St})$ characterizing the magnitude drag force and the coefficient of restitution $\al$ characterizing the inelasticity of collisions. The agreement for the reduced shear rate [see Fig.\ \ref{class} (a)] and the elements of the pressure tensor [see Figs.\ \ref{figPij} and \ref{figPii}] between Monte Carlo simulations and both theoretical solutions is excellent (especially in the case of nonlinear Grad's solution) for not too large value values of $\gamma^*$. As the magnitude of the friction coefficient increases the agreement between Grad's solution and simulations gets less good [cf. Fig.\ \ref{figPijalfa}], although being the discrepancies smaller than 6\%. This good performance of Grad's method has been also observed for monodisperse dry granular gases for Couette flow sustaining a uniform heat flux \cite{VSG10,VGS11,VSG13} and also in the case of granular binary mixtures under USF \cite{MG02,L04}. Regarding high velocity moments, we also obtain good agreement for the kurtosis $K$, since the BGK results compare very well with simulations for not too large values of $\gamma^*$ [cf. Fig.\ \ref{kurtosis}]. Finally, as expected, the BGK model reproduces very well the behavior of the marginal distributions $\varphi_x^{(+)}(c_x)$ and $\varphi_y^{(+)}(c_y)$ in the region of thermal velocities [see Figs.\ \ref{fxre} and \ref{fyre}], although they quantitatively disagree with simulations for higher velocities especially for strong collisional dissipation.

.

Finally, it is also important to remark that the objective of this work has been to set a non-linear hydrodynamic theory for the USF, state that as we know is necessarily non-Newtonian \cite{VSG10}, as a starting point for the deployment of a more comprehensive and systematic theory for more complex flows in this kind of system. In this context, we expect in the near future to extend the present results to other related flows such as the so-called LTu flows \cite{VSG10,VGS11} (i.e., the more general case of uniform but non-null heat flux) or to the more general class of Couette flows \cite{VSG13}. We want also to carry out further studies on the more realistic case of multicomponent granular suspensions where problems like segregation can be addressed. Work along these lines is underway.

\acknowledgments

The research has been supported by the Spanish Government through grant No. FIS2013-42840-P, partially financed by FEDER funds and by the Junta de Extremadura (Spain) through Grant No. GR15104.

\appendix
\section{Results from Grad's moment method. Rheological properties}
\label{appA}

In this Appendix we provide the approximate results obtained from Grad's moment method. First, we evaluate the collisional moment $\Lambda_{ij}$ defined in Eq.\ \eqref{3.14} by using Grad's approximation \eqref{3.1}. Before considering the trial distribution function \eqref{3.1}, the collision integral $\Lambda_{ij}$ can be written as
\beqa
\label{a1}
\Lambda_{ij}&=&m
\sigma^{d-1}\int\; \dd\mathbf{V}_1\dd\mathbf{V}_2\;f(\mathbf{V}_1)f(\mathbf{V}_2)\int \dd \widehat{\boldsymbol {\sigma}}
\Theta(\widehat{\boldsymbol {\sigma}}\cdot \mathbf{g})
\nonumber\\
&\times&(\widehat{\boldsymbol {\sigma }}\cdot \mathbf{g})\left(V_{1i}''V_{1j}''-V_{1j}V_{1j}\right),
\eeqa
where $\mathbf{g}=\mathbf{V}_1-\mathbf{V}_2$ is the relative velocity and
\begin{equation}
\label{a2}
\mathbf{V}_1''=\mathbf{V}_1-\frac{1+\alpha}{2}(\widehat{\boldsymbol {\sigma}}\cdot \mathbf{g})\widehat{\boldsymbol {\sigma }}.
\end{equation}
Using Eq.\ \eqref{a2}, Eq.\ \eqref{a1} becomes
\begin{eqnarray}
\label{a3}
\Lambda_{ij}&=&m\sigma^{d-1}\int\; \dd\mathbf{V}_1\dd\mathbf{V}_2\;f(\mathbf{V}_1)f(\mathbf{V}_2)
\nonumber\\
&\times  & \int \dd \widehat{\boldsymbol {\sigma}}\Theta(\widehat{\boldsymbol {\sigma}}\cdot \mathbf{g})\left[\left(\frac{1+\alpha}{2}\right)^2(\widehat{\boldsymbol {\sigma }}\cdot \mathbf{g})^3
\widehat{\sigma}_i\widehat{\sigma}_j\right.\nonumber\\
&-& \left.\frac{1+\alpha}{2}(\widehat{\boldsymbol {\sigma }}\cdot \mathbf{g})^2\left(
\widehat{\sigma}_j V_{1i}+\widehat{\sigma}_iV_{1j}\right)\right].
\end{eqnarray}
To perform the angular integrations, we need the results
\begin{equation}
\label{a4}
\int\; \dd \widehat{\boldsymbol {\sigma}}\Theta(\widehat{\boldsymbol {\sigma}}\cdot \mathbf{g})
(\widehat{\boldsymbol {\sigma}}\cdot \mathbf{g})^n=\beta_n g^n,
\end{equation}
\begin{equation}
\label{a5}
\int\; \dd \widehat{\boldsymbol {\sigma}}\Theta(\widehat{\boldsymbol {\sigma}}\cdot \mathbf{g})
(\widehat{\boldsymbol {\sigma}}\cdot \mathbf{g})^n \widehat{\boldsymbol {\sigma}}=\beta_{n+1} g^{n-1}\mathbf{g},
\end{equation}
\begin{equation}
\label{a6}
\int\; \dd \widehat{\boldsymbol {\sigma}}\Theta(\widehat{\boldsymbol {\sigma}}\cdot \mathbf{g})
(\widehat{\boldsymbol {\sigma}}\cdot \mathbf{g})^n \widehat{\boldsymbol {\sigma}}\widehat{\boldsymbol {\sigma}}=\frac{\beta_{n}}{n+d} g^{n-2}\left(n\mathbf{g}\mathbf{g}+g^2 {\cal \openone}\right),
\end{equation}
where ${\cal \openone}$ is the unit tensor and
\begin{equation}
\label{a7}
\beta_n=\pi^{(d-1)/2}\frac{\Gamma\left((n+1)/2\right)}
{\Gamma\left((n+d)/2\right)}.
\end{equation}
Taking into account these integrals, the integration over $\widehat{\boldsymbol {\sigma}}$ in Eq.\ \eqref{a3} yields
\beqa
\label{a8}
\Lambda_{ij}&=&-m\sigma^{d-1}\beta_3\frac{1+\alpha}{2}
\int \,\dd{\bf v}_{1}\,\int \,\dd{\bf v}_{2}f({\bf V}_{1})f({\bf V}_{2})g\nonumber\\
&\times&\left[g_i G_j+g_j G_i
+\frac{2d+3-3\alpha}{2(d+3)}g_i g_{j}-\frac{1+\alpha}{2(d+3)}g^2\delta_{ij}\right],
\nonumber\\
\eeqa
where ${\bf G}=({\bf V}_1+{\bf V}_2)/2$ is the center of mass velocity.

The expression \eqref{a8} is still \emph{exact}. However, to compute \eqref{a8} one has to replace the true $f(\mathbf{V})$ by its Grad's approximation \eqref{3.1}. The result is
\beq
\label{a9}
\Lambda_{ij}=-p n\sigma^{d-1}\sqrt{\frac{2T}{m}}(1+\al)\beta_3 I_{ij},
\eeq
where $I_{ij}$ is the dimensionless quantity
\beqa
\label{a10}
I_{ij}&=&\pi^{-d}\int \,\dd{\bf c}_{1}\,\int \,d{\bf c}_{2} e^{-(c_1^2+c_2^2)}\left[(c_{1\mu}c_{1\lambda}+
c_{2\mu}c_{2\lambda})\Pi_{\mu\lambda}^*\right.\nonumber\\
& & \left.+c_{1\lambda}c_{1\mu}c_{2\gamma}c_{2\nu}\Pi_{\mu\lambda}^*\Pi_{\gamma\nu}^*\right]
g^*\left[g_i^* G_j^*+g_j^* G_i^*\right.\nonumber\\
& & \left.
+\frac{2d+3-3\alpha}{2(d+3)}g_i^* g_{j}^*-\frac{1+\alpha}{2(d+3)}g^{*2}\delta_{ij}\right].
\eeqa
Here, $\mathbf{c}_i=\mathbf{v}_i/v_0$, $\mathbf{g}^*=\mathbf{g}/v_0$, $\mathbf{G}^*=\mathbf{G}/v_0$, $\Pi_{ij}^*=\Pi_{ij}/p$, and $v_0=\sqrt{2T/m}$ is the thermal velocity. The Gaussian integrals involved in the calculation of $I_{ij}$ can be easily computed by considering $\mathbf{g}^*$ and $\mathbf{G}^*$ as integration variables instead of $\mathbf{c}_1$ and $\mathbf{c}_2$. The corresponding integrals can be done quite efficiently by using a computer package of symbolic calculation. Here, we have used MATHEMATICA \cite{W96b}. The final expressions of $\Lambda_{ij}$  are given by Eq.\ \eqref{3.15} for $d=3$ and Eq.\ \eqref{3.16} for $d=2$.

Once the collisional moment $\Lambda_{ij}$ is known, the hierarchy \eqref{3.13} can be solved. According to the geometry of USF, the only non-zero elements of the pressure tensor are the off-diagonal element $P_{xy}=P_{yx}$ (shear stress) and the diagonal elements $P_{kk}$ ($k=x,y$ and also $z$, if $d=3$). The equations defining these elements (including the $zz$ element that would only raise if $d=3$) can be easily obtained from Eq.\ \eqref{3.13}. They are given by
\beq
\label{3.19}
2 a^* \Pi_{xy}^*+2\gamma^*(1+\Pi_{xx}^*)=\Lambda_{xx}^*,
\eeq
\beq
\label{3.19.1}
2\gamma^*(1+\Pi_{yy}^*)=\Lambda_{yy}^*,
\eeq
\beq
\label{3.20}
a^* (1 + \Pi_{yy}^*) +2\gamma^*\Pi_{xy}^*=\Lambda_{xy}^*,
\eeq
where  $\Lambda_{ij}^*\equiv \Lambda_{ij}/p \nu$. Note that in the physical case $d=3$, $\Pi_{zz}^*$ can be obtained from the constraint $\Pi_{zz}^*=-(\Pi_{xx}^*+\Pi_{yy}^*)$.

The solution to Eqs.\ \eqref{3.19}--\eqref{3.20} gives the elements $\Pi_{xx}^*$, $\Pi_{yy}^*$ and $\Pi_{xy}^*$ as functions of the reduced shear rate $a^*$. Note that $a^*$ is proportional to the square root of the (steady) temperature. In order to close the problem, we need an extra condition to express $a^*$ in terms of $\gamma^*$ and $\al$. This is provided by the energy balance equation \eqref{2.17}, whose dimensionless form is
\beq
\label{3.22}
-\frac{2}{d}\Pi_{xy}^* a^*=2\gamma^*+\zeta^*,
\eeq
where $\zeta^*$ is defined by Eqs.\ \eqref{3.17} and \eqref{3.18} for spheres and disks, respectively. Thus, the solution to Eqs.\ \eqref{3.19}--\eqref{3.22} provides the forms of $\Pi_{ij}^*$ in terms of the coefficient of restitution $\al$ and the (dimensionless) friction coefficient $\gamma^*$. On the other hand, given that the collisional moments $\Lambda_{ij}^*$ are nonlinear functions of $\Pi_{ij}^*$, Eqs.\ \eqref{3.19}--\eqref{3.22} must be solved numerically (nonlinear Grad's solution).

An analytical solution to Eqs.\ \eqref{3.19}--\eqref{3.22} can be easily obtained when one only considers linear terms to $\Pi_{ij}^*$ in the expressions \eqref{3.15} and \eqref{3.16} for $\Lambda_{ij}$. This was the approach considered by Sangani \emph{et al.} \cite{SGD04} to get the kinetic contributions to the pressure tensor at moderate densities. In this linear approximation (linear Grad's solution), the solution to Eqs.\ \eqref{3.19}--\eqref{3.22} can be written as
\beq
\label{3.23}
\Pi_{yy}^*=\Pi_{zz}^*=-\frac{\zeta_0^*+2\gamma^*}{\beta+\zeta_0^*+2\gamma^*}, \quad \Pi_{xx}^*=-(d-1)\Pi_{yy}^*,
\eeq
\beq
\label{3.24}
\Pi_{xy}^*=-\frac{\beta a^*}{(\beta+\zeta_0^*+2\gamma^*)^2},
\eeq
\beq
\label{3.24.1}
a^*=\sqrt{\frac{d (2\gamma^*+\zeta_0^*)}{2\beta}}(\beta+\zeta_0^*+2\gamma^*),
\eeq
where $\zeta_0^*\equiv \zeta_0/\nu$ is given by Eq.\ \eqref{2.18.1} and
\begin{equation}
\label{3.25}
\beta=\frac{1+\alpha}{2}\left[1-\frac{d-1}{2d}(1-\alpha)\right].
\end{equation}
In the dry granular case ($\gamma^*=0$), Eqs.\ \eqref{3.23}--\eqref{3.24.1} are consistent with previous results \cite{SGD04} obtained in the USF problem by using Grad's moment method. In addition, the expressions obtained by Sangani \emph{et al.} \cite{SGD04} agree with Eqs.\ \eqref{3.23}--\eqref{3.24.1} in the limit of dilute granular suspensions.

\section{Results from the BGK-like kinetic model}
\label{appB}

The exact results derived from the BGK-like kinetic model \eqref{3.26} are displayed in this Appendix. In terms of the dimensionless quantities $\widetilde{a}$, $\widetilde{\zeta}$ and $\widetilde{\epsilon}$, the BGK equation \eqref{3.26} can be rewritten as
\beq
\label{3.29}
\left(1-d\widetilde{\epsilon}-\widetilde{a}V_y \frac{\partial}{\partial V_x}-\widetilde{\epsilon}\mathbf{V}\cdot \frac{\partial}{\partial \mathbf{V}} \right)f(\mathbf{V})=f_\text{M}(\mathbf{V}).
\eeq
The hydrodynamic solution to Eq.\ \eqref{3.29} is
\beqa
\label{3.30}
f(\mathbf{V})&=&\left(1-d\widetilde{\epsilon}-\widetilde{a}V_y \frac{\partial}{\partial V_x}-\widetilde{\epsilon}\mathbf{V}\cdot \frac{\partial}{\partial \mathbf{V}} \right)^{-1}f_\text{M}(\mathbf{V})\nonumber\\
&=&
\int_0^{\infty}\; \dd t e^{-(1-d\widetilde{\epsilon})t}\;
e^{\widetilde{a}tV_y \frac{\partial}{\partial V_x}}\; e^{\widetilde{\epsilon}t\mathbf{V}\cdot \frac{\partial}{\partial \mathbf{V}}}
f_\text{M}(\mathbf{V}).\nonumber\\
\eeqa
The action of the velocity operators $e^{\widetilde{a}tV_y \frac{\partial}{\partial V_x}}$ and
$e^{\widetilde{\epsilon}t\mathbf{V}\cdot \frac{\partial}{\partial \mathbf{V}}}$ on an arbitrary function $g(\mathbf{V})$ is
\beq
\label{3.31}
e^{\widetilde{a}tV_y \frac{\partial}{\partial V_x}}g(\mathbf{V})=g\left(\mathbf{V}+\widetilde{a}tV_y \mathbf{\hat{x}}\right),
\eeq
\beq
\label{3.31.1}
e^{\widetilde{\epsilon}t\mathbf{V}\cdot \frac{\partial}{\partial \mathbf{V}}}g(\mathbf{V})=g\left(e^{\widetilde{\epsilon}t}\mathbf{V}\right).
\eeq
Taking into account these operators, the velocity distribution function $f$ can be written as
\beq
\label{3.32}
f(\mathbf{V})=n \left(\frac{m}{2T}\right)^{d/2}\varphi(\mathbf{c}),
\eeq
where $\mathbf{c}\equiv (m/2T)^{1/2}\mathbf{V}$ and the (scaled) velocity distribution function $\varphi(\mathbf{c})$ is
\beq
\label{3.33}
\varphi(\mathbf{c})=\pi^{-d/2}\int_0^{\infty}\; \dd t\;  e^{-(1-d\widetilde{\epsilon})t}\;
\exp \left[-e^{2\widetilde{\epsilon}t}\left(\mathbf{c}+t \widetilde{\mathbf{a}}\cdot \mathbf{c}\right)^2\right],
\eeq
where we have introduced the tensor $\widetilde{a}_{ij}=\widetilde{a}\delta_{ix}\delta_{jy}$.

Equations \eqref{3.32} and \eqref{3.33} provide the explicit form of the velocity distribution function in terms of the parameter space of the system. The knowledge of $f(\mathbf{V})$ allows us to evaluate its velocity moments. In order to accomplish it, it is convenient to introduce the general velocity moments
\beq
\label{3.34}
M_{k_1,k_2,k_3}=\int\; \dd \mathbf{V}\; V_x^{k_1} V_y^{k_2} V_z^{k_3} f(\mathbf{V}).
\eeq
The only nonvanishing moments correspond to even values of $k_1+k_2$ and $k_3$. Insertion of Eq.\ \eqref{3.33} yields
\beqa
\label{3.35}
M_{k_1,k_2,k_3}&=&n\left(\frac{2T}{m}\right)^{k/2}\pi^{-d/2} \int_0^{\infty}\; \dd t\; e^{-(1-d\widetilde{\epsilon})t}\;
\nonumber\\
& & \times
\int \dd \mathbf{c}\; c_x^{k_1}c_y^{k_2}c_z^{k_3}\; e^{\widetilde{a} t c_y\partial_{c_x}}\exp\left(-e^{2\widetilde{\epsilon}t}c^2\right)
\nonumber\\
&=&n\left(\frac{2T}{m}\right)^{k/2} \pi^{-d/2}\int_0^{\infty}\; \dd t\; e^{-(1+k\widetilde{\epsilon})t}\;
\nonumber\\
& & \times
\int \dd \mathbf{c}\; (c_x-\widetilde{a}t c_y)^{k_1}c_y^{k_2}c_z^{k_3} e^{-c^2},\nonumber\\
\eeqa
where $k=k_1+k_2+k_3$. It is now convenient to expand the term $(c_x-\widetilde{a}t c_y)^{k_1}$, so that
Eq.\ \eqref{3.35} becomes
\beqa
\label{3.36}
M_{k_1,k_2,k_3}&=&n\left(\frac{2T}{m}\right)^{k/2}
\sum_{q=0}^{k_1}\frac{k_1!}{q!(k_1-q)!}\langle c_x^{k_1-q}c_y^{k_2+q}c_z^{k_3}\rangle_{\text{L}}\nonumber\\
& & \times
\int_0^{\infty}\; \dd t\; (-\widetilde{a}t)^q e^{-(1+k\widetilde{\epsilon})t},
\eeqa
where
\beq
\label{3.37}
\langle c_x^{k_1}c_y^{k_2}c_z^{k_3}\rangle_{\text{L}}=\pi^{-3/2}\Gamma\left(\frac{k_1+1}{2}\right)
\Gamma\left(\frac{k_2+1}{2}\right)\Gamma\left(\frac{k_3+1}{2}\right)
\eeq
if $k_1$, $k_2$ and $k_3$ are even, being zero otherwise. Finally, after performing the $t$-integration in Eq.\ \eqref{3.36} one achieves the result
\beqa
\label{3.38}
M_{k_1,k_2,k_3}&=&n\left(\frac{2T}{m}\right)^{k/2}
\sum_{q=0}^{k_1}\frac{k_1!}{q!(k_1-q)!}(-\widetilde{a})^q \nonumber\\
& & \times (1+k\widetilde{\epsilon})^{-(1+q)}
\langle c_x^{k_1-q}c_y^{k_2+q}c_z^{k_3}\rangle_{\text{L}}.
\nonumber\\
\eeqa

In order to write more explicitly the form of the (scaled) distribution function $\varphi(\mathbf{V})$, we consider here a three-dimensional system ($d=3$). In this case, the distribution $\varphi$ can be written as
\beqa
\label{3.40.1}
\varphi(\mathbf{c})&=&\pi^{-3/2}\int_0^{\infty}\; \dd t\; e^{-(1-3\widetilde{\epsilon})t}\;
\nonumber\\
& & \times \exp\left[-e^{2\widetilde{\epsilon}t}(c_x+\widetilde{a}tc_y)^2-
e^{2\widetilde{\epsilon}t}c_y^2-e^{2\widetilde{\epsilon}t}c_z^2\right].
\nonumber\\
\eeqa
To illustrate the dependence of $\varphi$ on the parameter space of the problem, it is convenient
to introduce the following \emph{marginal} distributions:
\beq
\label{3.41}
\varphi_x^{(+)}(c_x)=\int_0^{\infty}\; \dd c_y\; \int_{-\infty}^{\infty}\; \dd c_z\; \varphi(\mathbf{c}),
\eeq
\beq
\label{3.43}
\varphi_y^{(+)}(c_y)=\int_0^{\infty}\; \dd c_x\; \int_{-\infty}^{\infty}\; \dd c_z\; \varphi(\mathbf{c}).
\eeq
Their explicit forms can be easily obtained from Eq.\ \eqref{3.40.1}:
\beqa
\label{phix}
\varphi_x^{(+)}(c_x)&=&\frac{1}{2\sqrt{\pi}}\int_0^{\infty}\;\dd t \; \frac{e^{-(1-\widetilde{\epsilon})t}}{\sqrt{1+\widetilde{a}^2t^2}}
\exp\left(-e^{2\widetilde{\epsilon}t}\frac{c_x^2}{1+\widetilde{a}^2t^2}\right)\nonumber\\
& & \times  \text{erfc}\left(
e^{\widetilde{\epsilon}t}\frac{\widetilde{a}tc_x}{\sqrt{1+\widetilde{a}^2t^2}}\right),
\eeqa
\beqa
\label{phiy}
\varphi_y^{(+)}(c_y)&=&\frac{1}{2\sqrt{\pi}}\int_0^{\infty}\;\dd t \; e^{-(1-\widetilde{\epsilon})t}
\exp\left(-e^{2\widetilde{\epsilon}t}c_y^2\right)\nonumber\\
& & \times \text{erfc}\left(
e^{\widetilde{\epsilon}t}\widetilde{a}tc_y\right).
\eeqa
In Eqs.\ \eqref{phix} and \eqref{phiy}, $\text{erfc}(x)$ is the complementary error function.

So far, $\chi$ has remained free. Henceforth, to agree with the results derived from linear Grad's solution, we will take $\chi=\beta$, where $\beta$ is defined by Eq.\ \eqref{3.25}.

\bibliography{suspension_shear}

\end{document}